%% file: ms-4axiv.tex
\documentclass{emulateapj}

\usepackage{graphicx}
\usepackage{epstopdf}
\usepackage{amsmath}

\def\simlt{\mathrel{\hbox{\rlap{\hbox{\lower4pt\hbox{$\sim$}}}\hbox{$<$}}}}
\def\simgt{\mathrel{\hbox{\rlap{\hbox{\lower4pt\hbox{$\sim$}}}\hbox{$>$}}}}

\def\ale{\mathrel{\hbox{\rlap{\hbox{\lower4pt\hbox{$\sim$}}}\hbox{$<$}}}}
\def\age{\mathrel{\hbox{\rlap{\hbox{\lower4pt\hbox{$\sim$}}}\hbox{$>$}}}}

\def\gs{\mathrel{\raise0.35ex\hbox{$\scriptstyle >$}\kern-0.6em
\lower0.40ex\hbox{{$\scriptstyle \sim$}}}}
\def\ls{\mathrel{\raise0.35ex\hbox{$\scriptstyle <$}\kern-0.6em
\lower0.40ex\hbox{{$\scriptstyle \sim$}}}}

\def\spose#1{\hbox to 0pt{#1\hss}}
\def\simlt{\mathrel{\spose{\lower 3pt\hbox{$\mathchar"218$}}
     \raise 2.0pt\hbox{$\mathchar"13C$}}}
\def\simgt{\mathrel{\spose{\lower 3pt\hbox{$\mathchar"218$}}
     \raise 2.0pt\hbox{$\mathchar"13E$}}}

\slugcomment{\it To be submitted to the Astrophysical Journal}

\shorttitle{SLSN Gaia16apd }
\shortauthors{Yan et al.}

\begin{document}

\title{Far-Ultraviolet to Near-Infrared Spectroscopy of A Nearby Hydrogen Poor Superluminous Supernova Gaia16apd}

\author{Lin Yan$^1$, R. Quimby$^2$,
A. Gal-Yam$^3$, P. Brown$^4$, N. Blagorodnova$^5$, E. O. Ofek$^3$, R. Lunnan$^5$, J. Cooke$^6$, S. B. Cenko$^{7,8}$, J. Jencson$^5$, M. Kasliwal$^5$}
\affil{$^1$ Caltech Optical Observatories \&\ Infrared Processing and Analysis Center, California Institute of Technology, Pasadena, CA91125, USA}
\affil{\it email: lyan@ipac.caltech.edu}
\affil{$^2$ Department of Astronomy, San Diego State University, CA92182, USA}
\affil{$^3$ Department of Particle Physics \&\ Astrophysics,  Weizmann Institute of Science, Rehovot 7610001, Israel}
\affil{$^4$ George P. and Cynthia Woods Mitchell Institute for Fundamental Physics \&\ Astronomy, Department of Physics and Astronomy, Texas A. \&\ M. University, 4242 TAMU, College Station,TX 77843, USA}
\affil{$^5$ Centre for Astrophysics \&\ Supercomputing, Swinburne University, Hawthorn VIC 3122, Australia}
\affil{$^6$ Department of Astronomy, California Institute of Technology, Pasadena, CA91125, USA}
\affil{$^7$ Astrophysics Science Division, NASA Goddard Space Flight Center, Mail Code 661, Greenbelt, MD 20771, USA}
\affil{$^8$ Joint Space-Science Institute, University of Maryland, College Park, MD 20742, USA}

\begin{abstract}
We report the first maximum-light far-Ultraviolet to near-infrared spectra (1000\AA\ $-$ 1.62$\mu$m, rest) of a hydrogen-poor superluminous supernova, Gaia16apd. At $z=0.1018$, it is the second closest and the UV brightest SLSN-I, with 17.4 magnitude in {\it Swift} UVW2 band at $-11$\,days pre-maximum. The coordinated observations with {\it HST}, Palomar and Keck were taken at $-2$ to $+25$\,days.  Assuming an exponential (or $t^2$) form, we derived the rise time of 33\,days and the peak bolometric luminosity of  $3\times10^{44}$\,erg\,s$^{-1}$.  At maximum, the photospheric temperature and velocity are 17,000\,K and 14,000\,km\,s$^{-1}$ respectively. The inferred radiative and kinetic energy are roughly $1\times10^{51}$ and $2\times10^{52}$\,erg. 
Gaia16apd is extremely UV luminous, emitting 50\%\ of its total luminosity at $1000 - 2500$\AA.  Compared to the UV spectra (normalized at 3100\AA) of well studied SN1992A (Ia), SN2011fe(Ia), SN1999em (IIP) and SN1993J (IIb), it has orders of magnitude more far-UV emission.  This excess is interpreted primarily as a result of weaker metal line blanketing due to much lower abundance of iron-group elements in the outer ejecta. Because these elements originate either from the natal metallicity of the star, or have been newly produced, our observation provides direct evidence that little of these freshly synthesized material, including $^{56}$Ni, was mixed into the outer ejecta, and the progenitor metallicity is likely sub-solar. This disfavors Pair-Instability Supernova (PISN) models with Helium core masses $\geq90M_\odot$, where substantial $^{56}$Ni material is produced. Higher photospheric temperature definitely contributes to the far-UV excess from Gaia16apd.  Comparing with Gaia16apd, we find PS1-11bam is also UV luminous.  

\end{abstract}


\section{Introduction}

UV spectra of supernovae (SNe) provide sensitive probes of the physical state of the ejecta and the environments, including element abundance, kinematic structure, density profile and ionization state \citep{Panagia2007, Bufano2009}.  Today, a little over 60 SNe have  early-time UV spectroscopy, which were taken by {\it HST}, {\it Swift}, {\it IUE} and {\it GALEX}.  Most of these sources are SN Ia, and a smaller fraction are core-collapsed events.  The UV flux from SNe Ia is generally a small fraction of the total emission due to metal line blanketing \citep{Pauldrach1996}, {\it i.e.} almost all of the UV photons initially produced in the inner layers of the ejecta  are absorbed by a forest of line transitions from single or doubly ionized iron-group elements.  These UV observations have instigated a flurry of theoretical studies which examined in detail the effect of ejecta abundance (or progenitor metallicity) \citep{Lentz2000,Mazzali2014}, reverse fluorescence (where ionized iron elements can convert photons from red to blue in the outer layers of ejecta) \citep{Mazzali2000}, ionization state (highly ionized iron will produce less UV absorption) \citep{Hoeflich1998, Sauer2008},  element mixing and velocity structures of the layers where UV photons are produced \citep{Hillebrandt2000}.   Type II SNe are generally luminous in UV at very early times (minutes to hours after explosion),  especially right after the shock wave produced by the core bounce has reached the stellar surface (shock breakout). After shock breakout and during the shock cooling phase,  with lower temperature and low ionization state,
metal line blanketing leading to suppression of UV continuum has been observed among SNe Ia and SN\,IIP \citep{Gal-Yam2008, Maguire2012, Brown2007, Dessart2008, Pritchard2014}.  However, UV spectra of SNe 1979C and 1980K from {\it IUE} have revealed an excess emission lines from highly ionized species such as N\,V, N\,III and Si\,IV, which were interpreted as emission from the interaction between ejecta and circumstellar medium (CSM) \citep{Panagia1980, Fransson1984}.

One of the new discoveries in the past ten years is SLSN \citep{Quimby2007, Barbary2009, Gal-Yam2012}, a rare class of supernovae which are difficult to explain by standard supernova models. An outstanding question in astronomy is what powers the energetic output from these events. 
SLSNe are broadly classified into two categories, ones without detectable hydrogen in their early-time spectra (SLSNe-I) and ones with hydrogen and/or helium emission (SLSNe-II, or luminous SNe\,IIn).  The characteristics of extreme peak luminosity and very long rise time scale implies that SLSNe may have massive progenitor stars, $>10M_\odot$ \citep{Nicholl2015}.  Optical spectroscopy of SLSNe-I has revealed features not commonly seen before, a series of six O\,II absorption troughs between 3200\,-\,4400\AA\ \citep[e.g.][]{Quimby2011}. These features are likely produced by O$^+$ ions with excitation potentials of 25\,eV, suggesting a very high temperature radiation field or energetic, non-thermal processes.  The relative strength of these O\,II absorption features in the early-time spectra can vary significantly from object to object. In some cases, only one or two features are visible \citep[i.e. SN2005ap;][]{Quimby2007, Yan2016}. This could make spectral classification difficult. One example is ASSASN-15lh whose early-time spectra have only one or two features from this O\,II series \citep{Dong2016}. This has resulted a very uncertain classification. The physical cause of the spectral difference among SLSNe-I is not yet understood. 

Existing UV spectra of superluminous supernovae (SLSNe) are rare, especially for low redshift events. A few rest-frame UV spectra reaching $\sim2000$\AA\  were obtained for SLSNe at $z>0.7$ from ground-based optical telescopes. 
Three characteristic absorption features are observed at 2200, 2500 and 2700\,\AA.  It is a topic of debate what exact ions produce these features, whether they are C\,II, Si\,II and Mg\,II as suggested by Quimby et al. (2011) or, C\,III/C\,II, C\,II, and C\,II/Mg\,II as proposed by Howell et al. (2013), or identified as C\,III/C\,II/Ti\,III, Ti\,III/C\,II/Si\,II and C\,II/Mg\,II blends by the modeling of Mazzali et al. (2016). 
The distinguishing power will need to come from the combined spectral data covering FUV, NUV and optical wavelengths.  

Only two FUV spectra exist for SLSNe-I. The first one is a noisy spectrum for PS1-11bam at $z=1.157$, taken with a ground-based optical telescope reaching down to the rest-frame 1300\,\AA.  The second one is from {\it HST} for ASASSN-15lh, which is a peculiar transient event whose true physical nature is still debated \citep{Dong2016, Leloudas2016,Brown2016,Godoy-Rivera2016}.  Clearly more early-time far-UV spectroscopy of SLSNe-I is needed.
Furthermore,  deeper transient surveys are capable of detecting SLSNe out to $z>4$ \citep{Cooke2012,Tanaka2013}. When these transient candidates are followed up with ground-based optical spectroscopy, the corresponding spectral features will be in the rest-frame NUV and FUV. Therefore, it is important that we can characterize the basic UV spectral properties of low-$z$ SLSNe.

In this paper, we report the first maximum-light ultraviolet spectra of a bright SLSN-I, Gaia16apd (SN2016eay), at $z=0.1018$.  Our far-UV to near-IR spectra cover a wide wavelength range from 1000\AA\ to 16,200\AA\ (rest-frame).  The observed spectral energy distribution has important implications for high redshift SLSN events. Throughout the paper, we adopt a $\Lambda$CDM cosmological model with
$\Omega_{\rm{M}}$\,=\,0.286, $\Omega_{\Lambda}$\,=\,0.714, and $H_0$\,=\,69.6\,$\rm{km}\rm{s}^{-1}\rm{Mpc}^{-1}$ \citep{Planck2015}.



\section{Observations}
\label{sec_obs}
\subsection{Our target}

Gaia16apd was first discovered as a transient event with $V$-band brightness of 17.3\,mag (AB) on 2016 May 16 by the Gaia Photometric Survey \citep{Gaia2016}. An optical spectrum taken on 2016 May 20 with the Nordic Optical Telescope \citep{Kangas2016} classified this event as a SLSN-I at $z=0.1018$ (473\,Mpc), making it the second closest SLSN-I among more than 60 discovered to date.  Gaia16apd is at the sky position of RA\,=\,12:02:51.71, DEC\,=\,+44:15:27.4 (J2000).
The first {\it Swift} observation on 2016 May 21 revealed that Gaia16apd is extremely bright in the UV (UVW2, 1928\,\AA) with a flux density of $2.54\times10^{-15}$\,erg\,s$^{-1}$\,cm$^{-2}$\AA$^{-1}$ (17.44 AB mag). 
Because of its early discovery and the extreme UV brightness revealed by the {\it Swift} data, we submitted a {\it HST} Director's Discretional Time (DDT) proposal (PID: 14516) and obtained early-time FUV and NUV spectra.  The {\it HST} COS and STIS UV spectroscopic observations were taken in three epochs, on 2016 June 2, 14, and 30 respectively.  
In addition,  optical and near-IR spectra were taken at Palomar and Keck around the same time as the {\it HST} data. 

The host galaxy is SDSS\,J120251.71+441527.4 with $u,g,r,i$ photometry of  22.13, 21.73, 21.76, and 21.19\,mag (AB) respectively, indicating a faint dwarf galaxy. The corresponding absolute magnitudes are $-16.11, -16.51, -16.48, -17.05$ (AB mag, K-corrected) respectively.  For comparison,  the \citet{Perley2016} study has shown the median values of $M_g$ and stellar mass $M_*$ of $-17.3$ AB mag and $2\times10^8M_\odot$ respectively based on a sample of 17 SLSN-I host galaxies. This suggests that the host of Gaia16apd is also a dwarf galaxy, consistent with other studies of SLSN-I host galaxies \citep{Perley2016,Lunnan2014,Leloudas2015}. 

Gaia16apd has a Galactic extinction E$(B - V)$ of 0.0132 \citep{Schlafly2011}.  Adopting R$_V$\,=\,$A_V/E(B-V)$\,=\,3.1 and Cardelli extinction law \citep{Cardelli1989}, we estimate the extinction at 1500\AA\ is only 0.1 magnitude, corresponding to less than 10\%\ increase in flux. Dust extinction correction is included in our analysis below.

\subsection{Light Curves, Explosion and Peak Dates}
\label{sec_phot}
\begin{figure*}[!ht]
\plotone{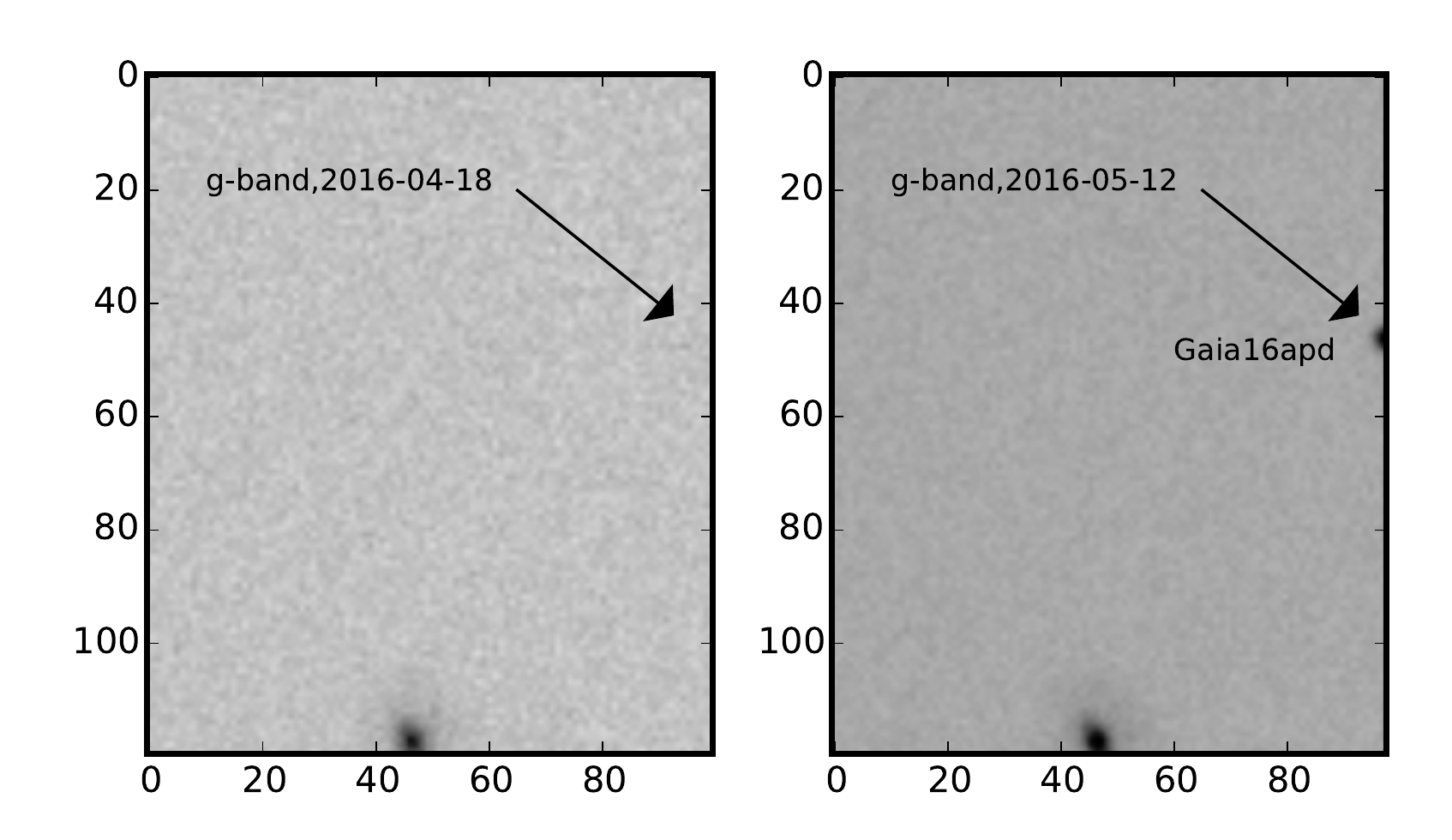}
\caption{The early time $g$-band images taken by Palomar 48inch telescope on 2016-05-12 and 2016-04-18. The axes are in pixels with a pixel scale of 1.01$^{''}$/pixel.  These data allow us to narrow down the explosion date.
 \label{p48image}}
\end{figure*}

The field containing Gaia16apd was also observed by the Palomar Transient Factory twice in May and 21 times in April 2016. Unfortunately, Gaia16apd fell on the edge of detector 0 in the images taken on 2016-05-12 and 2016-04-18, as shown in Figure~\ref{p48image}.  However, we do have enough pixels to perform Point-Spread-Function (PSF) fitting, and obtained a $g$-band magnitude of 17.3$\pm0.2$ (AB) on MJD = 57520.319~day (2016-05-12). The systematic error in this measurement is very large due to missing  pixels.  We used the PTFIDE software to carry  out the forced photometry on the reference subtracted images \citep{Masci2017,Ofek2012}. No pre-explosion activities were detected between 2012 and 2016-05-12 to a $3\sigma$ limit of 21 (AB mag), the sensitivity of a single exposure.  

The 3$\sigma$ limit on 2016-04-18 (MJD = 57496.275) is 21.0~mag (AB). 
From 2016-04-01 to 2016-04-18, there are a total of 21 $g$-band images from PTF.  Stacking the 9 images taken between 2016-04-18 and 2016-04-14 (MJD=57494.2$\pm$2\,days), the 3$\sigma$ limit is 21.3~mag (AB).  And coadding all of these images, we obtained 3$\sigma$ limit of 22.1~mag (AB). This upper limit covers the MJD range of $57487.7 \pm 8.5$\,days.  We note that because PTF has over several hundreds of $g$-band images taken between 2012 and 2016, we are able to make a good reference image. The stacking is done on the reference subtracted images, thus the derived magnitude limits are for the supernova only with the host light subtracted.

Gaia16apd was observed by the Ultraviolet and Optical Telescope \citep[UVOT;][]{Roming2005} on board of the {\it Swift} observatory \citep{Gehrels2004}, starting on 2016-05-21, an interval of $2-3$ days, over a month until 2016-06-23. This was the total amount of the approved time from {\it Swift}.  We followed the data reduction outlined for the {\it Swift} Optical Ultraviolet Supernova Archive \citep{Brown2014} and the {\it Swift} photometry calibration is based on \citet{Poole2008} and \citet{Breeveld2010}. 
Table~\ref{tab_phot} lists all of the broad-band photometry included in this paper. 

\begin{figure*}[!ht]
\plotone{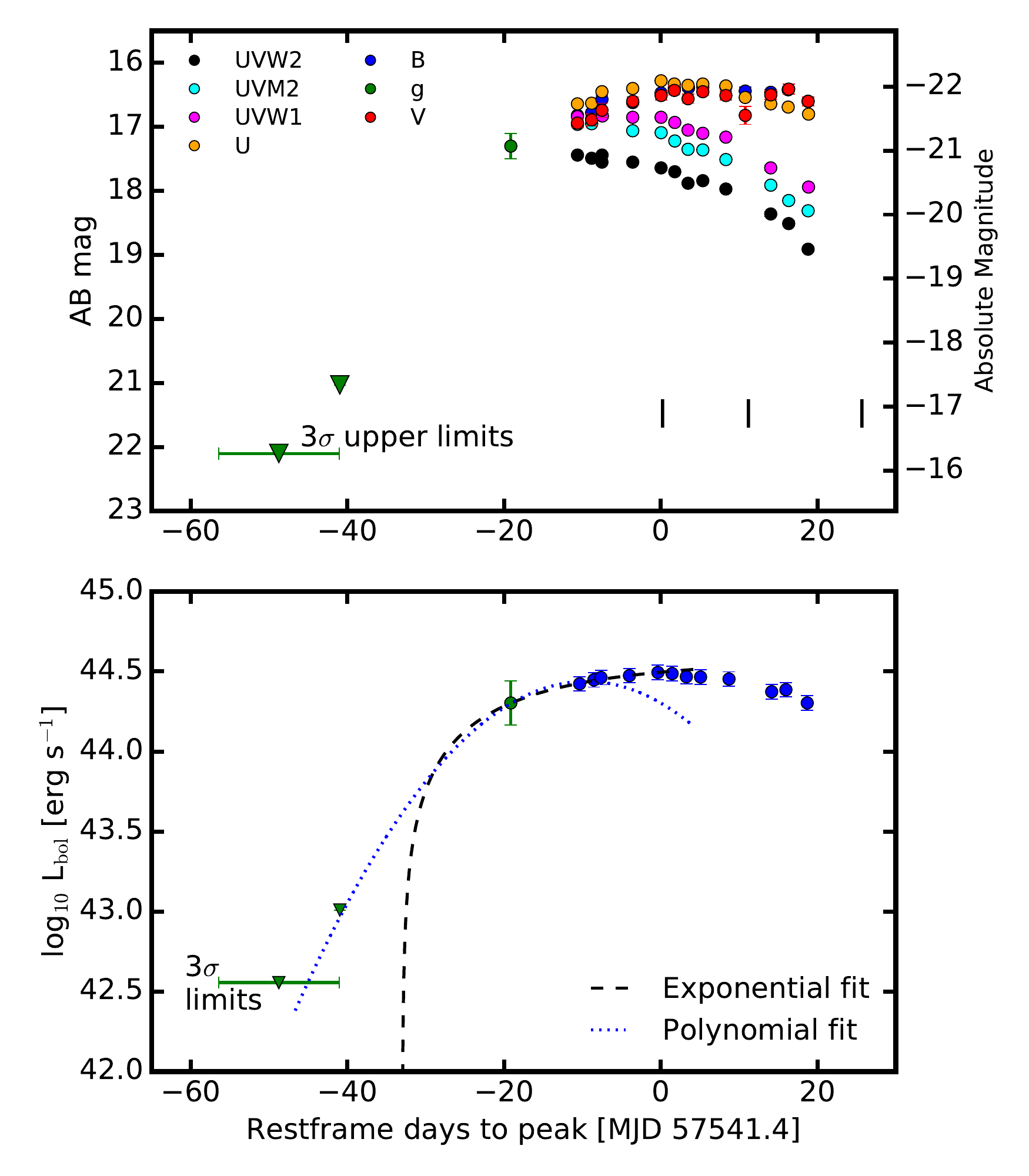}
\caption{The upper panel of this plot shows the early-time monochromatic light curves in various bands, including the limit from Palomar 48inch telescope.  The black vertical bars at the bottom of the upper panel mark the dates when the {\it HST} spectra were taken.
The bottom panel is the calculated bolometric light (data points) based on blackbody fits to the broad band photometries. The black dashed line is the exponential fit to the rising portion of the light curve. The dotted line is a 2nd-order polynomial fit to the early-time data. This curve sets the lower limit to the explosion date assuming the slowest rising rate.
\label{lc1}}
\end{figure*}

Figure~\ref{lc1} illustrates the monochromatic light curves (LC) in six {\it Swift} bands plus optical $g$-band.  All of the magnitudes here are the total magnitudes without any host galaxy subtraction.  It is worth noting that Gaia16apd is extra-ordinarily bright in {\it Swift} UV bands. As shown in Figure~\ref{lc1},  the highest, observed flux in the UVW2 (1928\AA) band is measured as $2.6\times10^{-15}$\,erg\,s$^{-1}$\,cm$^{-2}$ (17.5 AB mag) at the first epoch (2016-05-21, MJD=57529.7\,days), $-11$\,days before the bolometric peak date.  The UV fluxes of Gaia16apd declined only slightly (0.3\,mag) between $-11$ and 0\,days, when the {\it HST} UV spectroscopy was taken.

Bolometric luminosities (bottom panel) are the integrals of the blackbody fits to multi-band photometry. For the early times with only $g$-band data, we adopt blackbody temperatures from extrapolation of the multi-band estimates and scale the blackbody curve to match the observed $g$-band. 
One useful parameter is the rise time,  $t_{rise} = t_{peak} - t_{exp}$, which is directly related to photon diffusion time scale $t_{diff}$ and ejecta mass estimates (see below).  We have $t_{peak} = 57541.4$\,days. To determine $t_{exp}$, we use several different methods. One is to assume that early LC following a functional form, for example, an exponential form $L = L_{peak} (1.0 - \exp^{t_{exp} - t \over t_c})$  \citep{Ofek2014}, or a power-law form, $L \propto t^2$ like SNe\,Ia.  Although some studies claim that double-peak LCs could be prevalent among SLSNe-I, there are not a lot of concrete observational evidence supporting this hypothesis \citep{Nicholl2015}.  Since Gaia16apd does not have very early time photometry,  we adopt the assumption of smoothly rising, single peak profile.  With functional fitting, one may naively take the time as the explosion date when luminosity equals zero. Although mathematically this is correct, in practice, this method over-estimates $t_{rise}$. The reason is that just before and after explosion, light curve could be much steeper than the assumed exponential or power-law forms.  Here,  we defined the explosion date as when $L \sim 10^5L_\odot$, luminosity of a massive progenitor (hot blue supergiant).  The fitting using these two functional forms gives the similar result with $t_{exp} \sim 57505.3$\,days and $t_{rise} \sim 33$\,days, shown as dashed line in Figure~\ref{lc1}. 
This is shorter than that of other SLSNe-I published in the literature, although most of these have very few early-time data \citep{Nicholl2015,de Cia2016}.  Instead of assuming LC following certain functional forms, we use the flux limits to constrain the explosion date, shown as blue dotted line in Figure~\ref{lc1}. This yields $t_{exp} \sim 57461.9$\,days and $t_{rise}$ of 72\,days. The true value of $t_{rise}$ is likely to be smaller than this.

\subsection{Spectroscopy: from FUV to Near-IR}
\label{sec_spec}

Gaia16apd has an extensive spectroscopy dataset at early times. Table~\ref{tab_spec} summarizes all of the data, covering FUV, NUV, optical and near-IR.   

The {\it HST} DDT program was approved for a total of 5 orbits, with 3, 1 and 1 orbits for the observations taken on UT 2016-06-02 05:08:27, 
2016-06-14 04:47:38 and 2016-06-30 03:39:54 UT respectively.  Table~\ref{tab_HST} summarizes the observation parameters and the key features of the data.  The salient point is that the COS FUV spectra cover $1118-2251$\AA\ and the STIS/NUV data are from $1570-3180$\AA\ respectively. The COS and STIS spectra shown below are the reduced products from the {\it HST} archive. 

The optical spectra were taken with the Double Beam SPectrograph \citep[DBSP;][]{Oke1982} on the 200\,inch telescope at Palomar Observatory (P200), the Low-Resolution Imaging Spectrometer \citep[LRIS;][]{Oke1995} on the Keck telescope.  The near-infrared spectra in the $J$ and $H$ bands were taken with the Multi-Object Spectrometer For Infra-Red Exploration on the Keck telescope \citep[MOSFIRE;][]{McLean2012}. Table~\ref{tab_spec} lists the epoch, observatory and instrument, spectral coverage and resolution for each dataset.  Optical spectroscopy data are reduced with the software written by E. Bellm (DBSP) and D. Perley (LRIS).  The $J$ and $H$-band data was processed by the MOSFIRE Data Reduction Pipeline \footnote{See http://www2.keck.hawaii.edu/inst/mosfire/drp.html}.  An A0V-type star was observed immediately preceding Gaia16apd and its spectra are used to correct telluric absorption as well as flux calibration.

\section{Results}
\label{sec_result}

\subsection{Physical parameters derived from the LC}
Based on the bolometric light curve, the radiative energy emitted over the rest-frame 40\,days is $(7\pm0.7)\times10^{50}$\,erg.  Adopting the photon diffusion approximation and assuming $t_{diff} = t_{rise}$, we have $t_{rise} = \sqrt{2f\kappa M_{ej}/(c\times v_{ej})}$, where $f = {9 \over {4\pi^3}}$ and $\kappa$ is mass opacity \citep{Arnett1996, Padmanabhan2000}.  Here we assume $\kappa = 0.1$\,cm$^2$g$^{-1}$ commonly adopted for ejecta without H and He. It could be as high as 0.2\,cm$^2$g$^{-1}$ for a fully ionized H-poor medium \citep{Arnett1982}.   
As shown below, the ejecta velocity $v_{ej}$ can be measured from the optical spectra, and is roughly 14,000\,km\,s$^{-1}$.  Using the above equation, we have $M_{ej} = {t_{rise}^2 \,c\, v_{ej} \over 2f\, \kappa}$, thus, the estimated ejecta mass $M_{ej}$ is $12M_\odot$, with the upper limit of $57M_\odot$. This sets the minimum mass of the progenitor star for Gaia16apd.  The estimated kinetic energy is roughly $2\times10^{52}$\,erg. 
We note that this method of ejecta mass estimate is extremely crude because the assumption of $t_{diff} = t_{rise}$ could be far off for power sources other than $^{56}$Ni.  As discussed in detail in \citet{Nicholl2015},  the true diffusion time scale tends to be longer than $t_{rise}$ when central power sources are not $^{56}$Ni. This means our ejecta mass is under-estimated, and considered as only a lower limit. The proper estimate of diffusion time scale is to use a parametric fitting to the full light curve, including an assumed heating function.

\subsection{UV spectra of Gaia16apd}
\begin{figure*}[!ht]
\plotone{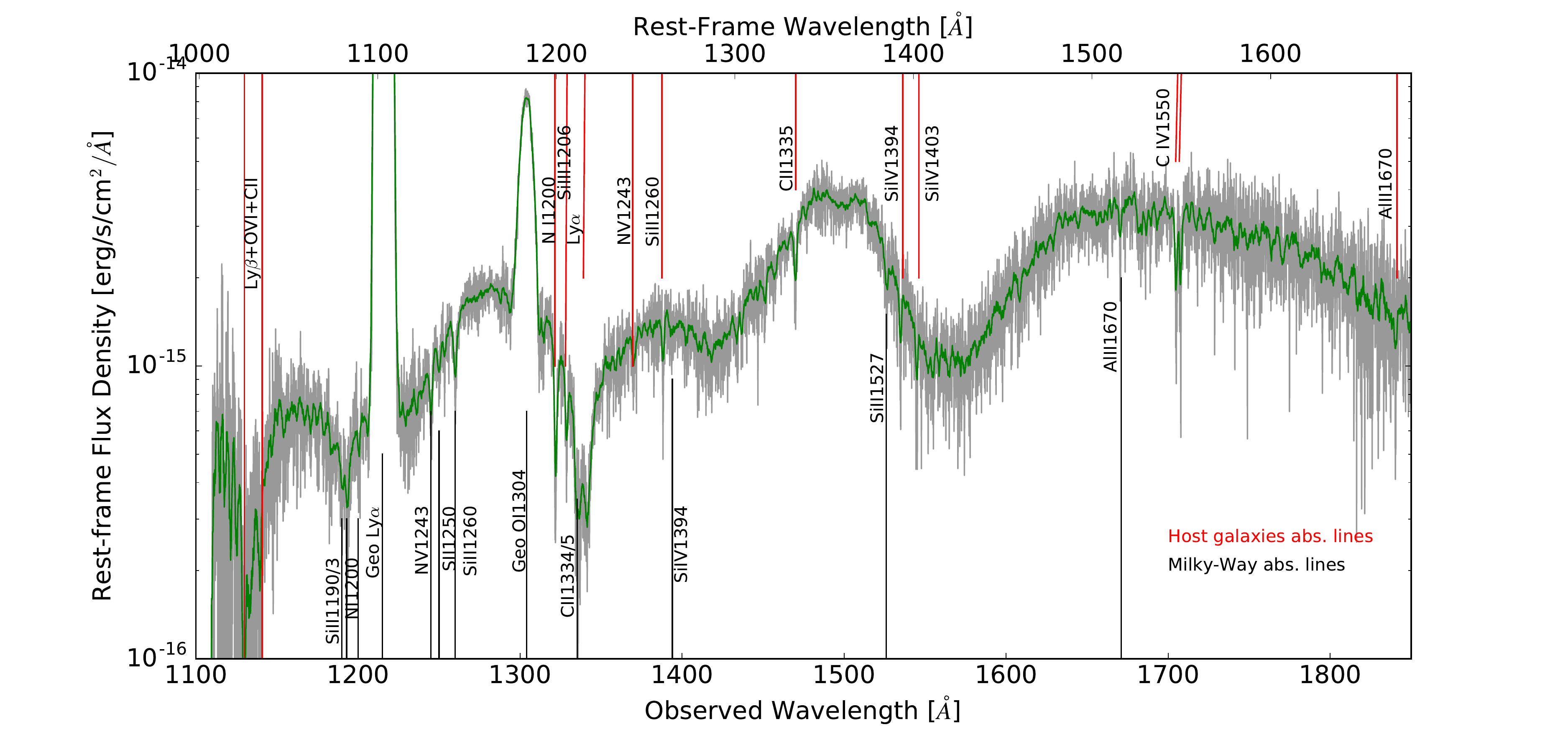}
\caption{The first far-UV spectrum of Gaia16apd (SLSN-I) taken with {\it HST/COS} 
on 2016-06-02, +0\,day relative to the peak date.  The grey spectrum is the original data and the green line is the smoothed version. The features from the Milky Way are marked with black vertical lines, and the ones from the host galaxy of Gaia16apd are marked with red vertical lines. The Y-axis is for the rest-frame flux density in erg/s/cm$^2$/\AA, and the X-axis is marked with both the observed (bottom) and the rest-frame (top) wavelength. 
\label{UVspec_absid}}
\end{figure*}

\begin{figure*}[!ht]
\plotone{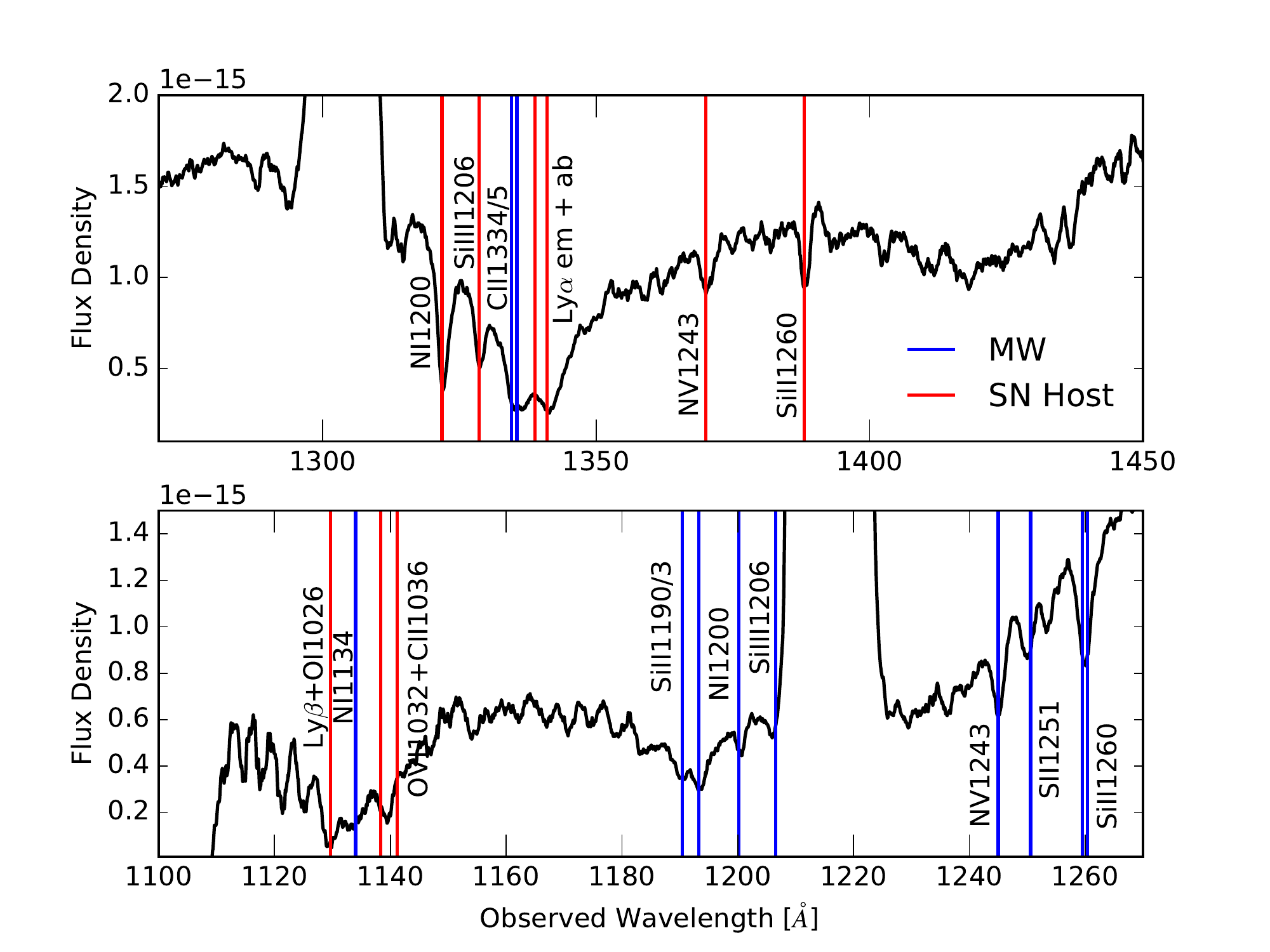}
\caption{Zoom-in around the three strong, blended absorption features in the far-UV spectra. We show that these three features are not due to the supernova Gaia16apd. The blue vertical lines mark the absorption lines due to the Milky Way, and the red vertical lines mark the lines from the host galaxy.
\label{uvzoom}}
\end{figure*}

Figure~\ref{UVspec_absid}  presents the first far-UV spectrum of a hydrogen-poor SLSN at a phase of 0\,days relative to the peak date (grey -- original resolution; green -- smoothed). We used a simple boxcar smoothing algorithm from Astropy\footnote{http://www.astropy.org}. Two prominent emission lines are geocoronal Ly$\alpha$\,1216\,\AA\ and O\,I\,1302\,\AA\ from the upper terrestrial atmosphere. Many narrow absorption lines are present, from both our Galaxy (marked in black lines) and the host galaxy of Gaia16apd at $z=0.1018$ (marked in red lines). There are three strong absorption features at the observed wavelengths of 1125, 1190 and 1340\,\AA, which are not related to the supernova and have been identified as blended absorption lines from the Milky Way and the host galaxy. 
Figure~\ref{uvzoom} presents the zoom-in version of these three features, with line identifications marked. It is worth noting that the host galaxy of Gaia16apd produces both damped Ly$\alpha$ absorption, as suggested by the line profile. We also see a weak Ly$\alpha$ emission line from the host as well as Ly$\beta$ absorption. This suggests that the SLSN is likely near the inner or backside of the host galaxy in project.  Although not required by this paper, the host galaxy extinction would be necessary for other studies which require accurate UV luminosities.  The wavelength of the host Ly$\alpha$ emission line is consistent with that of the host H$\beta$ and H$\alpha$ emission lines observed in the optical spectra. 

In the analyses below, we remove the geocoronal emission lines and the three strong absorption features in order to focus on the broad spectral features due to the supernova Gaia16apd.

\subsection{Full spectral energy distributions (SED) near the peak}
\begin{figure*}[!ht]
\plotone{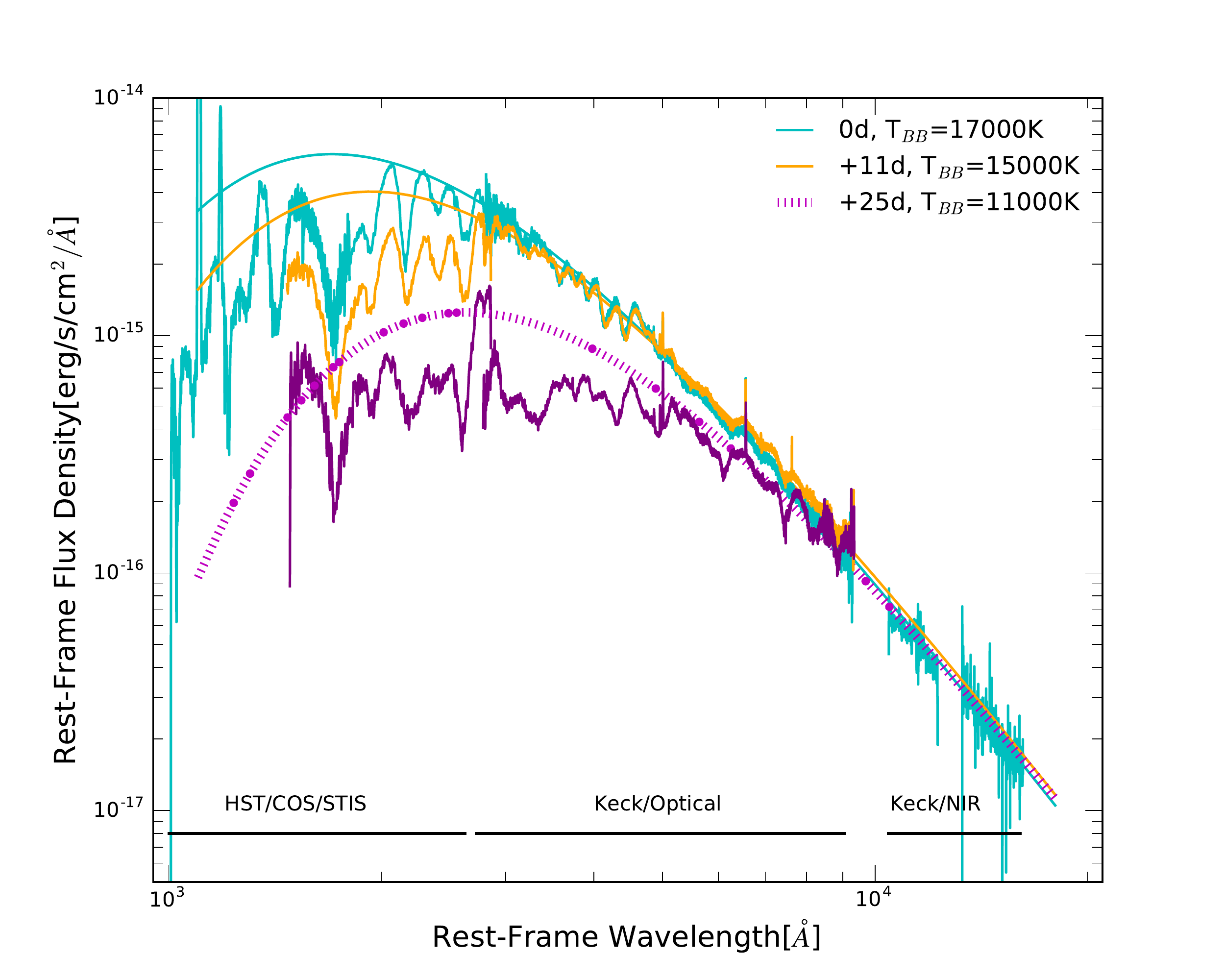}
\caption{We present all of the early-time spectra taken for Gaia16apd, covering the far-UV to the near-IR. The first
epoch (0\,day relative to the peak, rest-frame) spectra are in cyan, the second epoch (+11\,days) in orange, and the third (+25\,days) in purple. At the bottom of the figure, we mark the telescopes and instruments used for the observations.
\label{fullspec}}
\end{figure*}

One of the main results in this paper is the measurement of the spectral energy distribution of a hydrogen-poor SLSN, extending from the rest-frame far-UV 1000\,\AA\ to the near-IR 16,200\,\AA, shown in Figure~\ref{fullspec}.  This is the first early-time SED for a SLSN-I covering such a wide wavelength range. The SEDs are plotted for three epochs, +0, +11 and +25\,days relative to the peak date, determined by the {\it HST} UV observations. The corresponding optical spectra were chosen to closely match the phases of the UV data, see Table~\ref{tab_spec} for details.  At the maximum light, we need to stitch several spectra together in order to get the full SED. The {\it HST} UV spectra and {\it Swift} photometry were taken on 2016-06-02, whereas the closest optical spectrum was taken on 2016-05-31. We convolve the {\it Swift} broad band filters with the UV and optical spectra. Comparing the calculated magnitudes with that of measured from the {\it Swift} images, we found that {\it HST} flux calibration is consistent with that of {\it Swift} with differences less than 9\%, and the optical spectrum taken on 2016-05-31 needs to be scaled down by 11\%\ ({\it i.e.} multiplying the optical spectrum by a factor of 0.89). All colors are consistent without any corrections. This small flux correction now matches the optical spectrum to the UV spectra at the correct phase, 2016-06-02. Similar cross checking is also carried out for the second epoch when the {\it Swift} photometry is available.

All of the spectra, except NIR $J$ and $H$ band spectra, are plotted as they are, and the $J$ and $H$ spectra are shifted up by {\it multiplying} the original data by a factor of $3.1$ and $1.45$ respectively.  The smaller of these two factors is within the expected calibration uncertainty, and the large scaling factor is because the proper calibration data was missing, and the previous night calibration observation was used for the $J$-band data. Overall, the SED at the pre-peak phase can be fit by a black-body with temperature of 17,000\,K. At wavelengths short-ward of 2000\,\AA, the SEDs deviate from a blackbody form, with fewer photons coming through than the blackbody prediction.  This is due to some blended line absorption and also a small amount of metal line blanketing effect. As we argue in the later sections, line blanketing in Gaia16apd is orders of magnitude weaker than that of normal SNe\,Ia.  This reflects both the absence of newly synthesized heavy elements as well as intrinsically low metal abundance of the progenitor star (see detailed discussion below). 

The time evolution of the {\it HST} UV spectra can be summarized as follows.  As shown in Figure~\ref{fullspec}, the blackbody temperatures between the maximum light and $+25$\,day are falling rapidly from 17,000K to 11,000K.  
After our paper was submitted for publication, \citet{Nicholl2017} examined the time evolution of the UV spectral features.  They found that the equivalent widths of the UV absorption features have become slightly larger with time for Gaia16apd. For this calculation, blackbody fits are adopted as the assumed continua.

\begin{figure*}[!ht]
\plotone{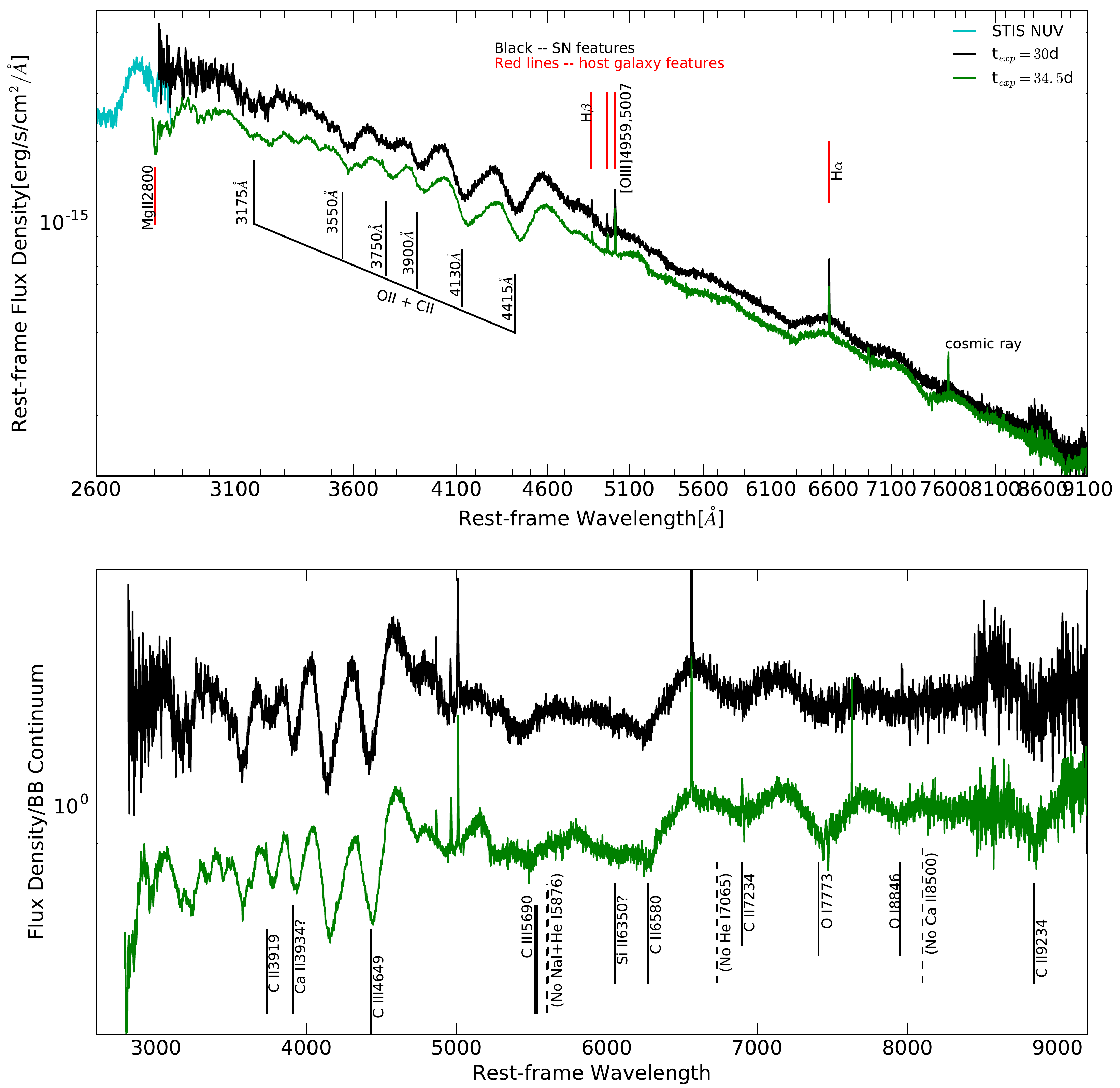}
\caption{The two optical spectra taken at the time closest to the first {\it HST} UV spectra are presented. The bottom panel shows the spectra with the continuum removed to high-light the rich set of absorption features detected in Gaia16apd.  Dashed vertical lines indicate the line transitions which are expected but not detected in the spectra. 
\label{optspec}}
\end{figure*}

We have collected multi-epoch optical spectra for Gaia16apd. The complete analysis of this dataset will be presented in a separate paper. Here we briefly discuss a couple of salient features observed in the early-time optical and near-IR spectra.  The first epoch {\it HST} UV spectra were taken on 2016-06-02 ($t_{peak}=0$\,day). The two optical spectra near that time were taken on 2016-05-31 ($t_{peak}=-2$\,days) and 2016-06-07 ($t_{peak}=4.5$\,days) with the P200 and the Keck telescopes respectively.  Figure~\ref{optspec} presents these two spectra in both their original form (top panel) as well as the spectra with the blackbody continuum removed (bottom).  In the top panel, we marked the well-known six O\,II absorption series \citep{Quimby2016}. The narrow emission lines are [O\,III]5007\AA, H$\beta$ and H$\alpha$ from the host galaxy. These features, in combination with the host galaxy lines in the {\it HST} spectra, give a precise redshift of 0.1018.  Using the minimum of the absorption feature O\,I\,7773, we measure the ejecta velocity of $\sim14000$\,km\,s$^{-1}$. 

In the bottom panel, we mainly focus on C and O absorption features. We identify C\,II6580, 7234, 9234\AA, C\,III4649 (blended with O\,II series) and C\,III5690\AA\, and O\,I7773, 8446\AA.  {\it These C\,II and C\,III features are very rarely identified in SLSNe-I except in one case, SN2015bn, where a possible C\,II was identified in \citet{Nicholl2016}}.  Here all the features are marked with 14,000\,km\,s$^{-1}$ blue-shift.  We also marked the positions of two He\,I features in order to confirm its absence in Gaia16apd.
C\,II7234\AA\ was detected in SN\,Ic 2007gr \citep{Valenti2008}, which is thought to be a carbon-rich SN\,Ic.  

\begin{figure*}
\plotone{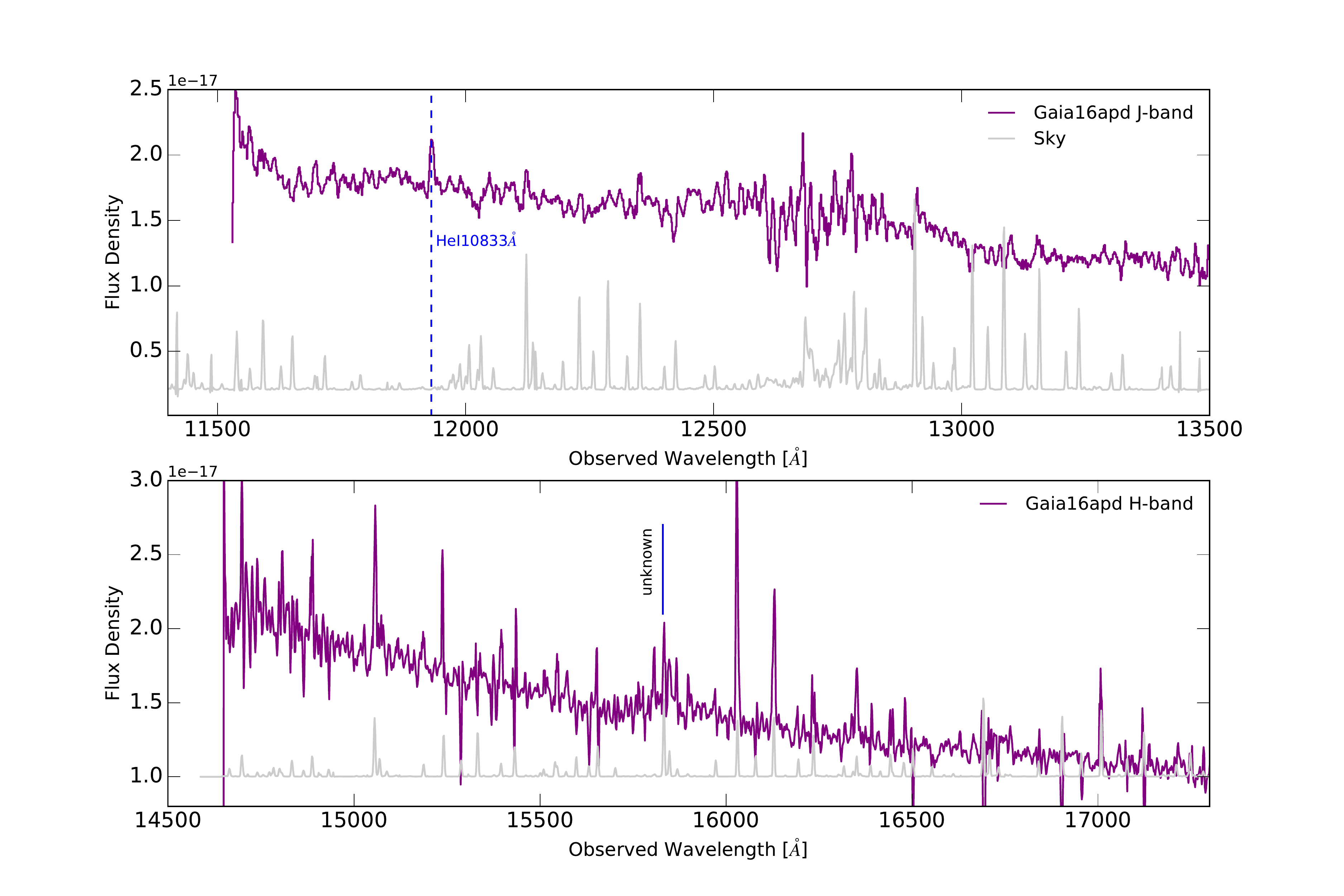}
\caption{The near-IR $J$ and $H$-band spectra are presented in the observed frame. In each panel, the $J$ and $H$ sky spectra taken with the same instrument (MOSFIRE) at the Keck telescope are plotted in light grey to indicate which regions of the Gaia16apd spectra are free of OH lines. 
\label{nirspec}}
\end{figure*}

Finally, the near-IR $J$ and $H$ spectra show mostly continua. Figure~\ref{nirspec} displays these two spectra in two panels. One significant spectral line is He\,I emission line at 1.0833$\mu$m.  We believe this He\,I emission is from the host galaxy, in accord with other narrow nebular emission lines such as Ly$\beta$, Ly$\alpha$, H$\beta$ and H$\alpha$, typical features from low luminosity dwarf galaxies. The lack of He\,I absorption supports the conclusion that Gaia16apd does not have any He in its optical and near-IR spectra. In general, He\,I\,10833\AA\ is rarely present in the spectra of SLSNe-I. The reported two cases are SN2012il \citep{Inserra2013} and SN2015bn \citep{Nicholl2016}, where the line identification for SN2012il is very uncertain. 
The second feature is a broad feature at 15830\,\AA\ (observed frame) in the $H$-band spectrum. This feature is in emission and fairly weak, corresponds to the rest-frame 14093\,\AA. Careful examination of the 2D spectra has confirmed the reality of this feature. However, its physical identification is still a mystery to us.   It is very puzzling that if this feature is associated with Gaia16apd, why it is in emission at the phase of maximum light since most SN spectral features are broad, blended absorption features.  If this broad feature were 
Hydrogen Brackett 14-4 transition at the rest-frame 15884.9\,\AA, the corresponding observed wavelength centroid would have been much
redder, at 17502\,\AA.
We carefully checked the telluric correction procedures to ensure this feature is not an artifact introduced during the removal of the telluric absorption feature between 15700 and 15800\,\AA. Additionally, this feature is not due to the removal of the nearby Hydrogen Brackett 4-14 absorption feature in the spectrum of the A0V standard used for flux calibrations.

\subsection{Far-UV excess emission from Gaia16apd}
\begin{figure*}
\plotone{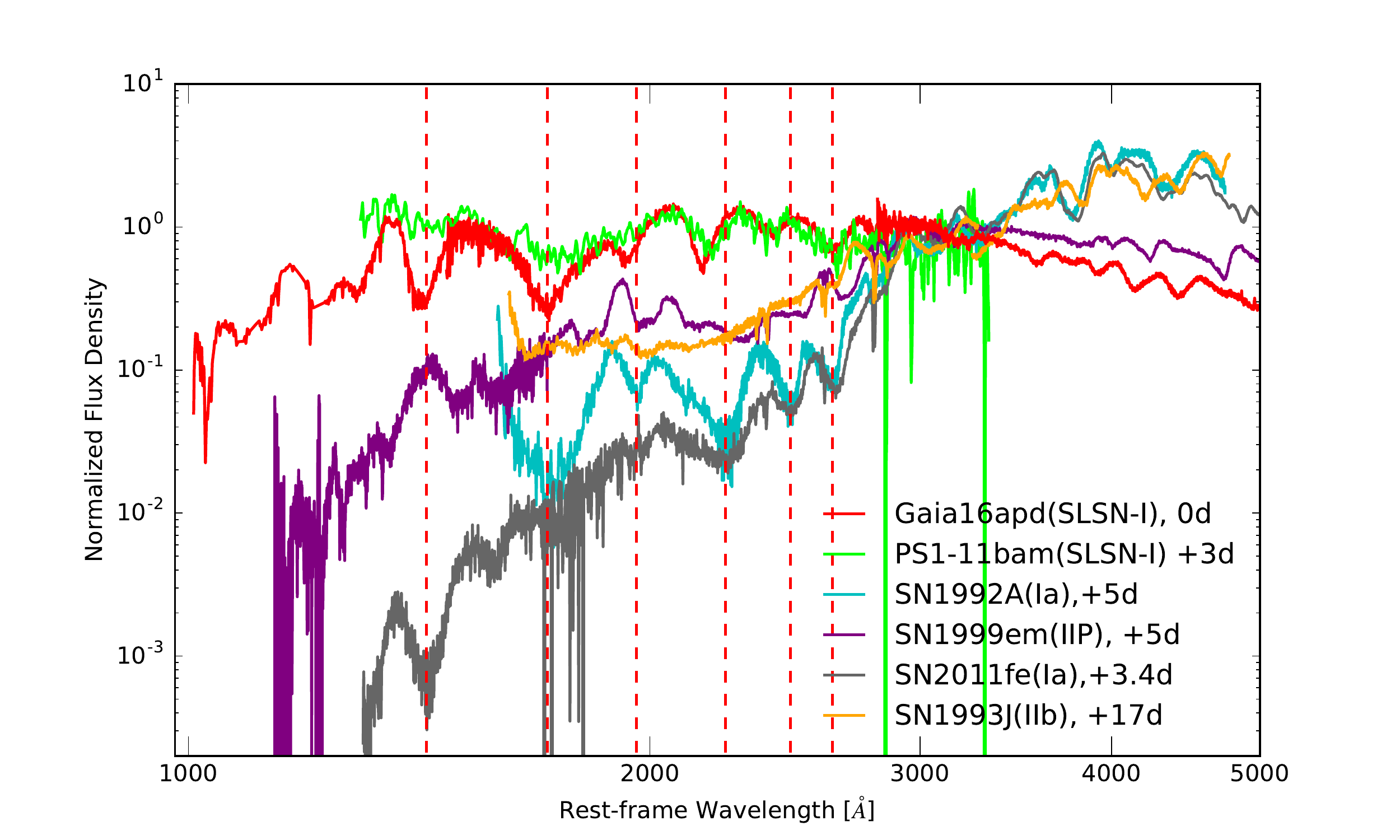}
\caption{We compared the normalized (at 3100\,\AA) UV spectra at similar early-phases between Gaia16apd (SLSN-I, red) and the lower luminosity SNe, including SN2011fe(Ia, grey), SN1992A(Ia, cyan), SN1999em (IIP, purple), SN1993J (IIb, orange).  The red dashed vertical lines mark the similar spectral features in both strengths and profiles in Gaia16apd and SN1992A.
\label{uvcompare}}
\end{figure*}

Figure~\ref{uvcompare} presents the UV spectrum of Gaia16apd at +0\,days, in comparison with that of other SNe with UV spectra normalized at 3100\AA. It is apparent that Gaia16apd is extra-ordinarily luminous in far-UV, and emits 50\%\ of its total luminosity at wavelength $<2500$\,\AA,  far exceeding any other normal SN.  Figure~\ref{uvcompare} makes a comparison with the rest-frame UV spectrum of PS1-11bam, a SLSN-I at $z=1.566$ \citep{Berger2012}. Although its SNR is not very high, the rest-frame UV spectrum of PS1-11bam behaves similarly as that of Gaia16apd, with a high fraction of UV emission.
This far-UV excess from SLSNe-I is further underscored by the comparison with the {\it HST} UV spectra of normal SNe at early-times, including SN\,1992A (Ia)  \citep{Kirshner1993},  SN\,2011fe (Ia) \citep[one of the closest SNe\,Ia;][]{Nugent2011,Mazzali2014,Foley2013}, SN\,1999em (IIP) \citep{Baron2000, Hamuy2001} and SN\,1993J (IIb) \citep{Jeffery1994}.   

It is well known that SNe\,Ia have relatively low fluxes in far-UV spectra at maximum light \citep{Maguire2012}.  The explanation is that normal SNe\,Ia produce abundant Fe group elements, including $^{56}$Ni, which subsequently goes through
$\beta$-decay to $^{56}$Co ($\tau_{half} = 6.1$\,days), then $^{56}$Co to $^{56}$Fe ($\tau_{half}$ = 77.7\,days), releases $\gamma$-ray photons which power the observed optical emission. For example, for SNe\,Ia, the average ejected $^{56}$Ni mass is $\sim 0.6M_\odot$. Single or doubly ionized heavy ions are known to have hundreds and thousands of overlapping line transitions, which strongly absorb the UV photons. This so called line blanketing effect is the reason why the far-UV continuum of a SN\,Ia is substantially suppressed. 

In supernova ejecta, iron-peak elements come from two different channels. One is the intrinsic metal content of the progenitor star. The second source, and more important one, is from newly synthesized material during the explosion. 
The observed far-UV continuum excess in Gaia16apd provides solid evidence that its outer ejecta
must not have  much iron-group elements, including $^{56}$Ni.  Otherwise, metal line blanketing would be obvious. If Gaia16apd has any $^{56}$Ni, it must be in the inner ejecta, and furthermore, very little mixing happened during the explosion.  Late-time observations will be important to confirm this prediction.  

Generally, UV spectrum is thought to form in the outer region of the ejecta, where there should be a substantial amount of unburned material, directly related to the surface layer of a progenitor star.
The apparent lack of strong UV line blanketing in Gaia16apd also suggests that the metal abundance of its progenitor star is probably sub-solar.  For example, \citet{Lentz2000} calculated the metallicity effects in Non-LTE model atmospheres of SNe\,Ia. They found that the model UV spectra can increase fluxes at 1500\AA\ easily by a factor of 10 when varying metallicity from 1 solar to 1/10 solar (their Figure 2).  Their synthetic far-UV spectra at $3$ and $10Z_\odot$ show significant line blanketing.  

\begin{figure*}
\plotone{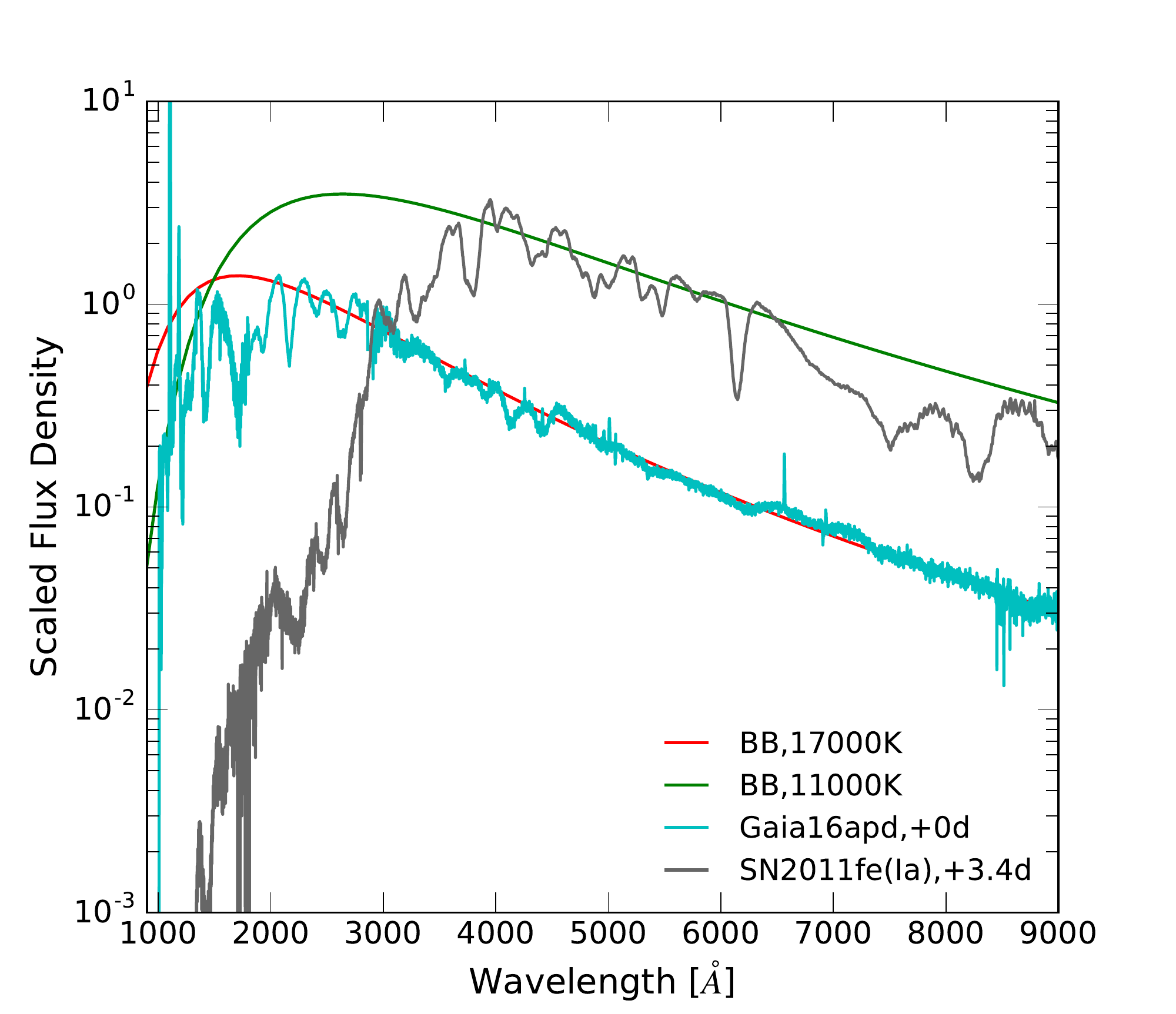}
\caption{This plot is to illustrate both temperature and line blanketing effects in Gaia16apd and SN2011fe. The line blanketing in Gaia16apd is much weaker than that of SN2011fe.  The spectra are normalized at 3000\,\AA.
\label{lineblk}}
\end{figure*}

Finally, SLSNe-I tend to have higher photospheric temperatures than that of normal SNe, for example, Gaia16apd has a $T_{BB} \sim 17,000$\,K, about a few thousands degree hotter than that of SNe\,Ia.  So how do we know which factor is the dominant source for the high UV flux in Gaia16apd: weak iron group line blanketing or high photospheric temperature? Figure~\ref{lineblk} plots the observed SEDs of Gaia16apd, SN2011fe together with their corresponding blackbody curves. The two spectra are normalized at 3000\,\AA. This figure shows clearly that line blanketing in SN2001fe is much stronger than that of Gaia16apd, and is the most important reason for its far-UV excess. Higher photospheric temperatures do shift overall SEDs toward shorter wavelengths, producing more UV photons. Hotter temperature can also keep more Iron group elements at higher ionization states, that contribute less to UV line blanketing. However, the ionization state in Gaia16apd may not be very high because of the detections of weak, low ionization lines such as C\,II6580, 7234, 9234\AA, and O\,I7773, 8446\AA\ in the optical region. 

We conclude that the primary reason for the observed far-UV excess in Gaia16apd is that its outer ejecta must have very little iron-group elements. This rules out the presence of newly synthesized $^{56}$Ni, unless it is in the inner region of the ejecta and without any mixing.  Furthermore, Our data suggests that the metal abundance of the progenitor star may be sub-solar. These results set very specific constraints on future explosion models.

The strong UV excess and low metal blanketing at maximum light suggest that Pair-Instability supernova (PISN) model may not work for  Gaia16apd.  Particularly PISN models with C$+$O core masses $\geq 90M_\odot$ synthesize a substantial amount of $^{56}$Ni \citep{Heger2002}.  The 33\,days rise time scale is too short compared to the PISN model predictions \citep{Kasen2011}, although our estimate has a quite large uncertainty. However, because newly synthesized $^{56}Ni$ tends to be in the inner parts of ejecta, without mixing, far-UV spectra of early-time LCs may show very little absorption by iron group elements.  To resolve this ambiguity, we need to follow up with Gaia16apd and its late-time decay slope should determine if  $^{56}Ni$ could be a  significant power source. 
In addition, PISN models with smaller core masses $64-90M_\odot$ produce very little $^{56}Ni$ \citep{Heger2002}. The main difficulty with this scenario is a lack of power source(s) for the optical light curves. Because PISN models do not produce any compact remnants (no neutron stars and no blackholes), power sources such as magnetars or fall-back blackhole accretion are not available.  One could argue that ejecta-CSM (H-poor) or H-poor shell-shell collisions, as predicted by Pulsational Pair Instability models \citep{Woosley2016}, could provide the energy source. However, the {\it Swift} XRT observations from +11.6 to $-53.6$\,days  yielded no detections, with 90\%\ confidence limits ranging from $6.5\times10^{-13}$ to $1.3\times10^{-13}$\,erg\,s$^{-1}$\,cm$^{-2}$ assuming a power law spectrum with a photon index of $+2$.  The stacked flux limit at 90\%\ confidence is $1.3\times10^{-14}$\,erg\,s$^{-1}$\,cm$^{-2}$, corresponding to the 0.3 $-$10\,Kev luminosity of $3.4\times10^{40}$\,erg\,s$^{-1}$. {\it This limit is about $3\times10^{-4}$ of $L_{bol}^{peak}$, an order of magnitude smaller than the predicted by CSM interaction models \citep{Svirski2012,Ofek2013}.}  This suggests that interaction is probably not important for this event.  Our spectra also rule out the \citet{Moriya2010} models, where energetic core collapse of a $\sim40M_\odot$ C$+$O core could explain the LCs of some SLSNe-I, but these models produce a large amount of ejected $^{56}$Ni.

It is worth noting that the PS1-11bam spectrum is similar to that of Gaia16apd with a high UV continuum. This may be an indication that UV luminous SLSNe-I like Gaia16apd could be more common than previously known.  However,  PTF12dam is a counter example to Gaia16apd. It is a SLSN-I at $z=0.107$, a similar redshift as that of Gaia16apd. At $-11$\,days, Gaia16apd has a UVW2 - V ([1928\AA] - [5430\AA]) color of 17.44 - 16.99 = 0.45\,mag (AB). In contrast, PTF12dam has a UVW2 - V color of $\sim2$\,mag (AB) at $-20$\,days \citep{Vreeswijk2017,Nicholl2013}.  This is much redder than that of Gaia16apd, implying that PTF12dam has much less UV flux relative to optical than that of Gaia16apd.  

After our paper was posted, \citet{Nicholl2017} has taken our {\it HST} data and made a comparison with the rest-frame UV spectra at peak or pre-peak phases from other 9 SLSNe-I, including PTF12dam, SN2015bn, SN2010gx, PTF09atu, iPTF13ajg, PS1-10ky, SCP-06F6, SNLS-06D4eu, SNLS-07D2b and PS1-11bam (see their Figure 3).  Similar to what we found, \citet{Nicholl2017} concluded that Gaia16apd is indeed a very unique event. Out of a total 10 SLSNe-I, including Gaia16apd, only 3 (33\%) have UV excess at $1000 - 3000\AA$ similar to that seen in Gaia16apd.  
Most SLSNe-I (77\%) have much less UV emission at $1000 - 3000\AA$, {\it i.e.} their UV spectral curves are significantly below that of Gaia16apd.  The same comparison was also made at $\sim +20$ to $+30$\,days post-peak, and shows that all SLSNe-I, including Gaia16apd have similar blackbody temperatures. 

In Figure~\ref{uvcompare}, red dashed lines mark six significant UV absorption features. At the first glance, there appears to be some similarity between SN\,Ia SN2011fe and SLSN-I Gaia16apd, if the blue shifts of some of the features are due to relatively higher photospheric velocity.  The question is if these six features are from the same ions.  Although detailed spectral identifications based on numerical calculations properly counting all elements and transitions are beyond the scope of this paper, we argue that these features can not come from the same physical transitions. These six features in SN\,Ia SN2011fe were identified mostly due to iron group elements, blends of Si\,II+Co\,II+Fe\,II (1st), Fe\,II+Ni\,II+Co\,II (4th), Fe\,II+Co\,II (5th) and Fe\,II+Mg\,II (6th) based on models by \citet{Mazzali2014}.  These same transitions can not be responsible for all of the UV features from Gaia16apd because it would contradict with the fact that its bright UV emission indicates very little metal line blanketing.  As discussed in \S\ref{sec_uvid},  the UV features from Gaia16apd are likely due to intermediate elements, such as Si\,III, C\,III, C\,II and Mg\,II \citep{Mazzali2016}.  Our current theoretical modelings of UV spectra of SLSNe-I are very limited.  We can not rule out some of these UV features from SLSNe-I and SNe\,Ia could indeed come from the same physical transitions.

\subsection{New UV spectral features \label{sec_uvid}}
\begin{figure*}[!ht]
\plotone{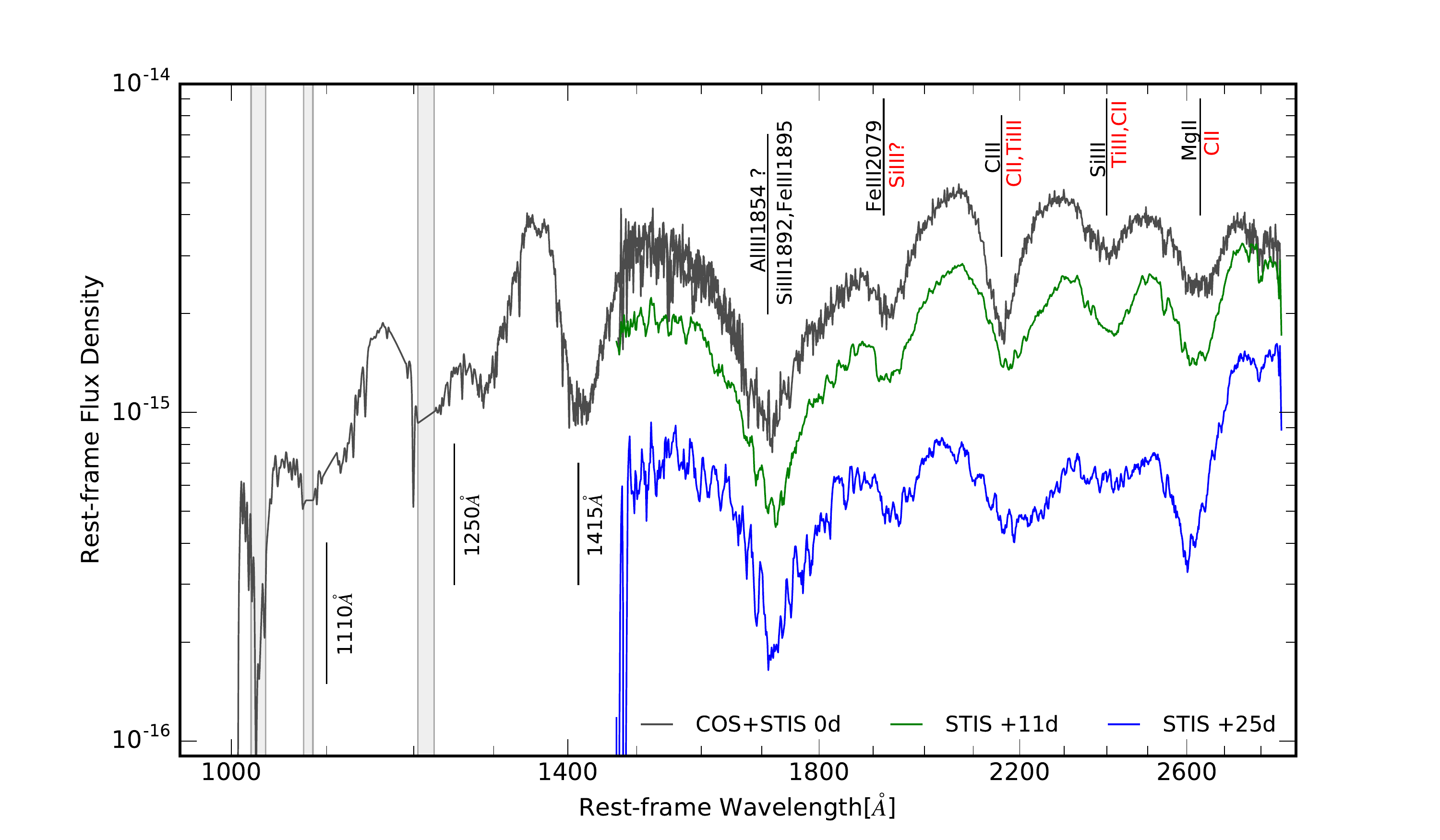}
\caption{This plot presents the one epoch of far-UV spectrum from {\it HST/COS} at 0\,days and the three epochs of near-UV spectra at 0, +11 and +25\,days relative to the peak. We mark the features we identified based the published papers as well as our own analysis.  Three new far-UV features are marked with the black vertical lines at 1110, 1250 and 1415\,\AA.  We used the grey vertical bars to mask the regions where deep absorption features are not associated with Gaia16apd, but due to the Milky Way and the host galaxy. 
\label{snuvspec}}
\end{figure*}

Figure~\ref{snuvspec} presents the far-UV and near-UV spectra taken at 0, +11 and 25\,days relative to optical flux peak using {\it HST} COS and STIS.  Here, we focus only on the features related to the supernova, with other features related to the Milky Way and the host galaxy removed or masked by the grey vertical bars.

\begin{figure*}
\plotone{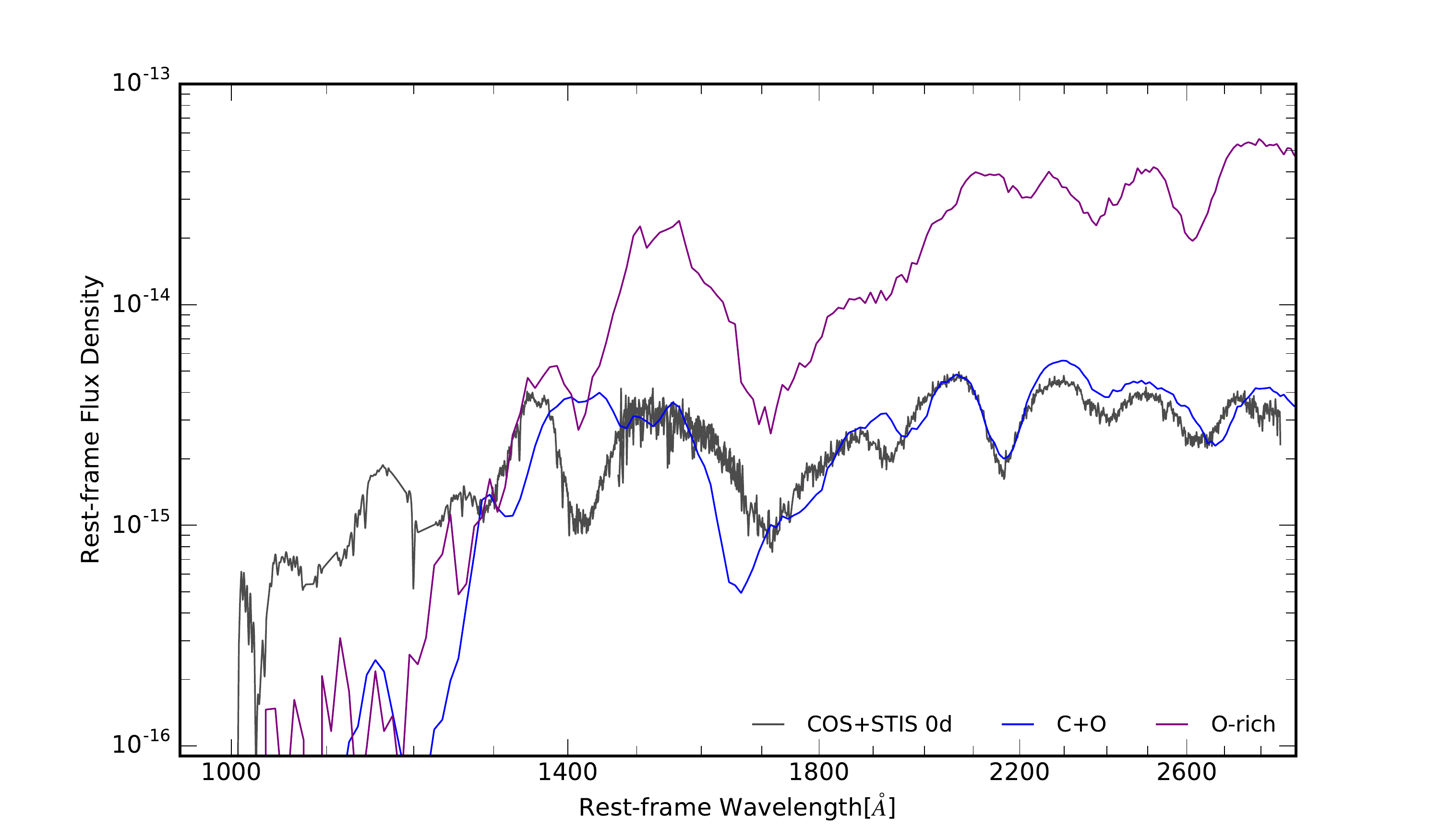}
\caption{This plot compares our UV spectra at 0\,day relative to the peak with two model spectra published in \citet{Howell2013} for two high-$z$ SLSNe-I. The data is in black, the C$+$O model spectrum is in blue, and the C$+$O with Oxygen-rich is in purple. For the details of the models, see the text.
\label{compare_kasen}}
\end{figure*}

The three broad features short-ward of 1500\,\AA\ are new, marked by the black vertical lines. We attempt to shed some light on the feature identifications by comparing with model spectra published for two high-$z$ SLSNe-I in \citet{Howell2013} in Figure~\ref{compare_kasen}. These model spectra were calculated for an envelope of an assumed composition above an inner boundary blackbody using 
the Monte Carlo radiative transfer code {SEDONA} \citep{Kasen2006}.  The normalization is determined by the assumed ejecta mass $M_{ej} = 5M_\odot$, kinetic energy $E_{kin} = 10^{52}$\,erg, peak bolometric luminosity $L_{peak} = 2.0\times10^{44}$\,erg\,s$^{-1}$,  and time since explosion $t = 25$\,days. These parameters are close to those of Gaia16apd, with time $t$ slightly earlier than that of our first UV spectrum (at $t = 30$\,days). 
The two model spectra in Figure~\ref{compare_kasen} are for two different compositions. The blue line is for  C$+$O model where all of the hydrogen and helium in the solar abundance was converted to equal parts of carbon and oxygen and the purple is for C$+$O model enhanced in Oxygen.  

Compared with these two model spectra, the UV spectrum of Gaia16apd clearly has a UV excess.  However, the model spectra seem to be able to crudely reproduce some of the observed features, particularly at $\lambda > 1600$\,\AA. Particularly the Oxygen-rich model spectrum (purple) seems to produce most spectral features, but fits poorly to the spectral slope.
The metallicities of the model spectra are clearly too high, under-predicting the far-UV fluxes below 1400\AA. It is worth noting that only the Oxygen-rich model (purple) seems to be able to explain the newly observed feature at 1400\,\AA, and the simple C$+$O model (blue) does not work at all.  Although it has been suggested that SLSNe-I are associated with massive C$+$O cores \citep{Quimby2011,Pastorello2010}, the far-UV spectra from Gaia16apd  now provide additional new insights on the properties of the C$+$O cores of stripped massive stars, which may produce hydrogen-poor superluminous supernovae.

\citet{Mazzali2016} has carried out a modeling of SLSN-I iPTF13ajg spectrum covering from $1800 - 6000$\AA.  The three prominent UV features at the rest-frame 2200, 2400, and 2700\AA\  (4th, 5th and 6th features marked in Figure~\ref{uvcompare} respectively) are modeled as blends of C\,III+C\,II+Ti\,III (4th), Si\,III+Ti\,III,\,C\,II (5th), and Mg\,II+C\,II (6th).  As shown in Figure~\ref{snuvspec}, the features at 1700 and 1950\AA\ (2nd and 3rd marked in Figure~\ref{uvcompare}) could be  Al\,III+Si\,III+Fe\,III (2nd feature) and Fe\,III+Si\,III (3rd) based on \citet{Mazzali2016} model.

In addition to comparing with the published calculations, we use syn++ code \citep{Thomas2011} to identify the potential ions producing the observed UV features.  The results from this exercise should be regarded as suggestive, and by no means a complete nor physically consistent modeling of the data.

\begin{figure*}
\plotone{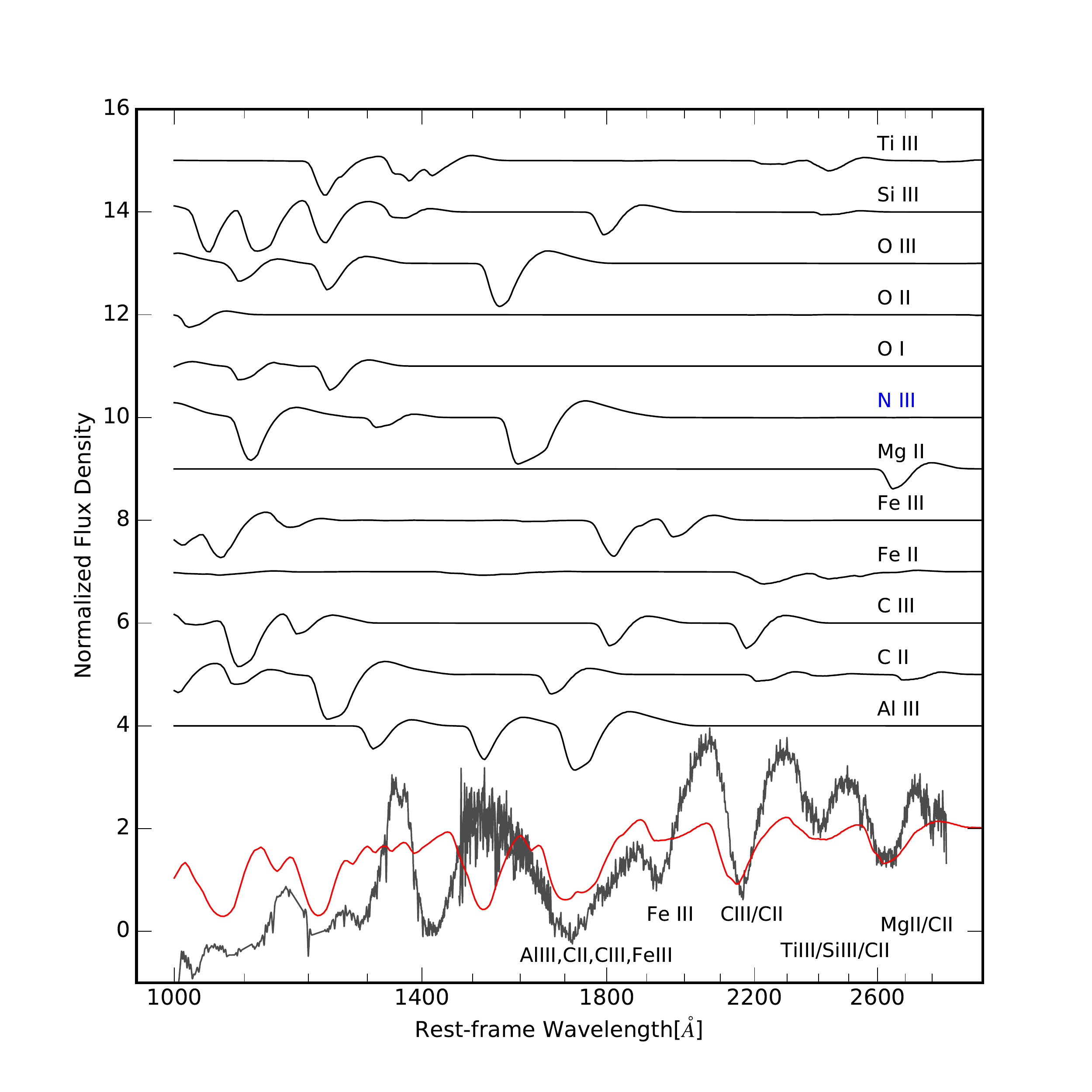}
\caption{This plot compares the Gaia16apd UV spectra at 0\,day with the synthetic spectra generated by syn++ \citep{Thomas2011}.  The spectra are normalized and shifted vertically for clear display.  The top 12 spectra were made for each corresponding ion, the bottom black spectrum is the observed data, and the red line is the synthetic spectrum including all of the ions except N\,III (labeled in blue). 
\label{compare}}
\end{figure*}

Figure~\ref{compare} shows the comparison between our data and the calculated spectrum for each individual ion, and the combined synthetic spectrum which including Al\,III, C\,II, C\,III, Fe\,II, Fe\,III, Mg\,II\, O\,I, O\,II, O\,III, Si\,III and Ti\,III. We exclude N\,III (marked as blue) from the combined spectrum because this ion produces a prominent feature near 1600\,\AA, which causes strong disagreement with the observed spectrum. We have also calculated spectra for MnII/III, CoII/III and CrII/III, all of these spectra have much weaker features in this region compared to that of FeII/III.  

Of the new features at $\lambda < 1800$\,\AA, the broad absorption at 1700\,\AA\ could be produced by Al\,III, C\,II, C\,III, Fe\,III. However, we note that including Al\,III also produces a strong feature at  1500\,\AA\ which is not present in our data.  This implies that even if Al\,III could contribute to the formation of 1700\,\AA\ feature, it may be small amount.  However, none of the ions listed above seems to have any feature at 1400\,\AA.  The spectrum from the O-rich C+O core model (Figure~\ref{compare_kasen}, purple) has a narrow absorption feature around 1400\,\AA. However, this feature was not identified in \citet{Howell2013}.  In addition, the wavelength region $\lambda < 1400$\AA\ was not the focus of their paper, and there is clearly too much absorption at the blue end.  Improved calculations are clearly needed. 
The same feature at 1400\AA\ is also seen in SN\,Ia SN2011fe (Figure~\ref{uvcompare}). \citet{Mazzali2014} paper tentatively identified it as Si\,II+Co\,II+Fe\,II (1st).  It is not clear if this same set ions is responsible for the 1400\AA\ feature in SLSN-I Gaia16apd because its bright UV emission rules out a lot of iron group elements in its outer layer ejecta.
We conclude that to fully understand the observed UV spectra, more theoretical work is needed.

Lastly, we note that the {\it HST} UV spectra of Gaia16apd are dramatically different from that of ASSASN-15lh \citep{Dong2016,Godoy-Rivera2016,Brown2016}.  Since Gaia16apd appears to be a typical SLSN-I, we conclude that ASSASN-15lh's UV spectra are not consistent with the earlier classification of being a SLSN-I. Other interpretations are discussed in literature \citep{Leloudas2016,Margutti2016}.

\subsection{High-$z$ SLSNe-I}

\begin{figure*}
\plotone{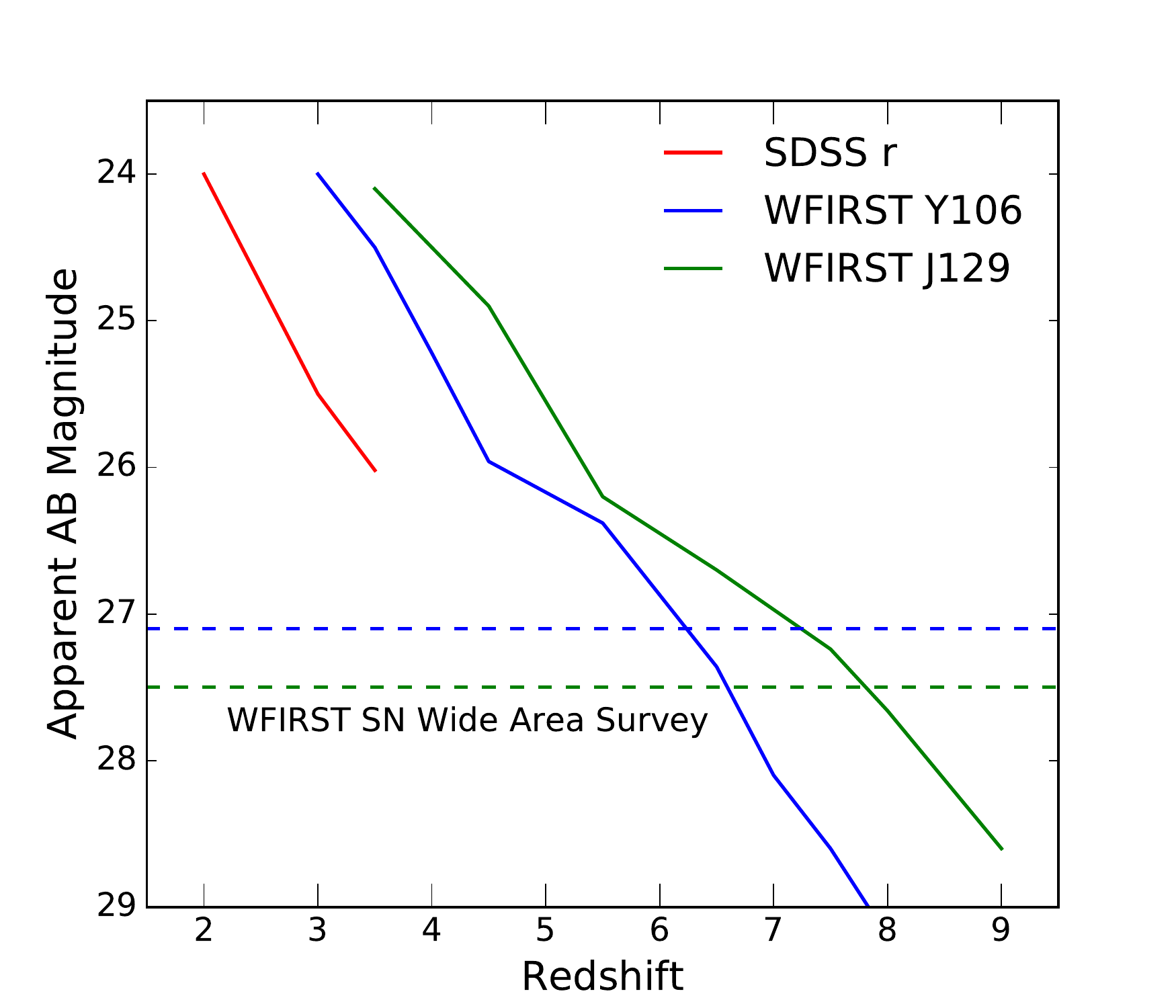}
\caption{ The estimated apparent AB magnitudes at maximum light for a SLSN-I as a function of redshifts, assuming the same intrinsic SED as we measured for Gaia16apd. \label{magz}}
\end{figure*}

At maximum light, Gaia16apd has a bolometric absolute magnitude of $-22.49$, corresponding to a bolometric luminosity of $3\times10^{44}$\,erg\,s$^{-1}$.  
If we assume that all high-$z$ SLSNe-I have the same peak absolute magnitude and UV SEDs similar to that of Gaia16apd at the maximum light, we can easily calculate the apparent magnitude as a function of redshift.  Comparing these peak apparent magnitudes with some of the future wide area, time-domain surveys, we can determine if these events at high-$z$ may be detectable.  Figure~\ref{magz} illustrates such a simple calculation, the estimated apparent magnitude in three possible bands, SDSS $r$ for intermediate redshifts, and WFIRST Y106W (0.927 $-$ 1.192$\mu$m) and J129W (1.131 $-$ 1.454$\mu$m) filters for high redshifts.  One of the main goals of WFIRST is Type Ia Supernova Survey (SN), which has two campaigns, each with 5 days cadence for 6 months \citep{Spergel2015}. These two campaigns are 2 years apart. The WFIRST supernova survey covers 27.4 sq.degree with 5$\sigma$ sensitivities of 27.1 in Y106W and 27.5 mag(AB) in J129W filters.  Therefore, it is promising that we may be able to detect these energetic events out to redshifts of 8.

However, a proper prediction calculation should take into account of peak luminosity and SED distributions. 
High-$z$ SLSNe-I must follow a distribution of peak luminosities as well as a range of UV SEDs.  So far observational selection of SLSNe does not have a set of strict criteria. They are commonly selected by peak absolute magnitudes brighter than $-20.5$ to $-21$, followed by spectral classification. Currently, the brightest SLSN-I is iPTF13ajg, with $L_{peak} \sim 6.3\times10^{44}$\,erg\,s$^{-1}$ ($-23.3$\,mag) \citep{Vreeswijk2014}, and the lower limit of an absolute magnitude of $-20.5$ corresponds to $L_{peak} \sim 5\times10^{43}$\,erg\,s$^{-1}$.  Therefore, Gaia16apd is on the brighter side of the peak luminosity distribution.  An unbiased peak luminosity distribution function will require future statistically complete samples.


In addition, as we discussed earlier, at peak or pre-peak phase, Gaia16apd has much more UV emission than most other SLSNe-I, and its blackbody temperature appears to be higher than that of most other SLSNe-I. This clearly illustrates that the UV SEDs of SLSNe-I are diverse, particularly at peak or pre-peak phases.  The real difficulty is to construct a statistically unbiased distribution function.  These required analyses are beyond the scope of this paper.

\section{Summary and Discussion}

The key results from this paper are following.  Gaia16apd is one of the closest SLSNe-I ever discovered, only at 473\,Mpc. Its proximity and the extreme UV brightness enable us to obtain the first maximum-light ultraviolet to near-infrared spectrum (1000\AA $-$1.62$\mu$m) of a SLSN-I.  The high SNR {\it HST} UV spectra at the maximum light revealed extremely luminous UV continuum,  
emitting 50\%\ of its total total luminosity at $1000 - 2500$\,\AA. This is in stark contrast to the UV spectra of other normal supernovae, such as SN2011fe (Ia), SN1992A (Ia), SN1999em (IIP) and SN1993J (IIb), where metal line blanketing of UV photons are significant.  

Gaia16apd also has early time optical images from the Palomar Transient Factory before it was triggered as a transient event by the Gaia Mission. Our analyses of the photometric light curves infer that Gaia16apd took 33\,days to reach its bolometric luminosity of $3\times10^{44}$\,erg\,s$^{-1}$. Its total radiative energy over the 60\,days since its discovery $E_{rad}$ is $1\times10^{51}$\,erg.  Assuming photon diffusion time scale to be the same as the rise time scale, the estimated ejecta mass is $12M_\odot$ with opacity $\kappa$\,=\,0.1\,cm$^2$\,g$^{-1}$.  With the photospheric velocity of $\sim$14,000\,km\,s$^{-1}$ measured from the optical spectra, we calculate the kinetic energy at the explosion $E_{kin} > 2\times10^{52}$\,erg.  This is a powerful event compared to normal SNe, but its kinetic energy is fairly typical for SLSNe. 

In ultraviolet wavelengths, iron-group elements have hundreds and thousands of lines.  Far-UV photons can be easily absorbed by these transitions and are very sensitive to the presence of these ions.  Heavy elements in supernovae come from two different sources. One is from freshly synthesized material. Another channel is the intrinsic metal abundance of the progenitor star of a supernova.   The observed far-UV excess in Gaia16apd has two implications: (1) its outer ejecta must not have much newly formed iron-peak elements, including $^{56}$Ni. If there is any $^{56}$Ni, it must be in the inner regions and there is very little mixing. (2) the progenitor of Gaia16apd  is a massive star whose metallicity is likely to be sub-solar.   We also argue that the high blackbody temperature at the maximum light may also contribute to the far-UV excess in Gaia16apd.  

Our result clearly rules out PISN models \citep{Heger2002} as well as energetic core collapse models \citep{Moriya2010}. Particularly the PISN models with progenitor masses $\geq 90M_\odot$ are definitely not viable because in this mass regime, a large amount of $^{56}$Ni material is produced.  The 33\,day rise time scale also argues against PISN models because it is quite shorter than model predictions \citep{Kasen2011}.  

Although the complete and reliable identification of the UV absorption features requires future detailed modelings, we made a tentative comparison with the published synthetic UV spectra (made available by D. Kasen). This comparison suggests that Gaia16apd may be an explosion of a massive O-rich C$+$O core with a sub-solar metal abundance.  Our high SNR UV spectra have revealed well detected absorption features, which should set constraints on the chemical composition, ionization state and temperature of the ejecta modeled by future theoretical studies.

Finally, we utilized our measured SED at the maximum light for Gaia16apd and estimated the apparent peak magnitudes in three different filters for various redshifts. We show that NASA future near-infrared space mission WFIRST may provide an opportunity to detect SLSNe-I out to redshift of 8.


\acknowledgments

We thank Daniel Kasen for very helpful discussion and for making his model spectra available.  Vikram Ravi is acknowledged for obtaining some spectra used in this paper. 
We are grateful to the {\it HST} staff for the prompt scheduling of these ToO observations.  We acknowledge ESA Gaia, DPAC and the Photometric Science Alerts Team\footnote{http://gsaweb.ast.cam.ac.uk.alerts}.
Based on observations made with the NASA/ESA Hubble Space Telescope, obtained from the Data Archive at the Space Telescope Science Institute, which is operated by the Association of Universities for Research in Astronomy, Inc., under NASA contract NAS 5-26555. M.M.K acknowledges support by the GROWTH project funded by the National Science Foundation under PIRE Grant No 1545949. E.O.O is incumbent of the Arye Dissentshik career development chair and is grateful for support bygrants from the Willner Family Leadership Institute Ilan Gluzman (Secaucus NJ), Israel Science Foundation, Minerva, and the I-Core program by the Israeli Committee for Planning and Budgeting and the Israel Science Foundation (ISF).  A.G.-Y. is supported by the EU/FP7 via ERC grant No. 307260, the Quantum Universe I-Core program by the Israeli Committee for planning and funding, and the ISF, Minerva and ISF grants, WIS-UK ``making connections'', and Kimmel and YeS awards.  
Some of the data presented herein were obtained at the W. M. Keck Observatory, which is operated as a scientific partnership among the California Institute of Technology, the University of California, and the National Aeronautics and Space Administration. The Observatory was made possible by the generous financial support of the W. M. Keck Foundation. The authors wish to recognize and acknowledge the very significant cultural role and reverence that the summit of Mauna Kea has always had within the indigenous Hawaiian community. We are most fortunate to have the opportunity to conduct observations from
this mountain.

{\it Facilities:} \facility{Swift}, \facility{HST (COS, STIS)}, \facility{Keck (LRIS, MOSFIRE)}, \facility{Palomar}.

{\it Software:} Astropy, SEDONA, syn++.

\begin{deluxetable}{cccc}
\tablecolumns{4}
\tablewidth{0pc}
\tabletypesize{\scriptsize}
\tablecaption{\label{tab_phot} Photometry$^a$ }
\tablehead{\colhead{MJD} & \colhead{Filter}  & \colhead{Mag} & \colhead{error} \\
\colhead{day} & & \colhead{mag} & \colhead{mag} }
 \startdata
\input{"./table_phot.tex"}
 \enddata
 \tablenotetext{a}{All magnitudes are in AB system.}
 \tablenotetext{b}{Errors with \nodata means the photometry is a 3$\sigma$ upper limit.}
\end{deluxetable}

\begin{deluxetable*}{ccccccc}
\tablecolumns{7}
\tablewidth{0pc}
\tabletypesize{\scriptsize}
\tablecaption{\label{tab_HST} {\it HST/UV} Spectroscopy Observation Log }
\tablehead{\colhead{Obs.UT} & \colhead{No.Orbits} & \colhead{Instrument} & Grating & \colhead{$\Delta \lambda$}  & \colhead{Spec Resolution} & \colhead{Obs.setup} \\
                                               &                                 &                                    &               & \colhead{\AA}   &   &  \\
 }
 \startdata
\input{"./table_HST.tex"}
 \enddata
 \tablenotetext{a}{COST/FUV data was taken using only Segment A.}
 \end{deluxetable*}
 
\begin{deluxetable*}{cccccc}
\tablecolumns{6}
\tablewidth{0pc}
\tabletypesize{\scriptsize}
\tablecaption{\label{tab_spec} Spectroscopy Observations of Gaia16apd }
\tablehead{\colhead{Date} & \colhead{MJD} & \colhead{exp. time} & \colhead{$\Delta \lambda$}  & \colhead{Spec Resolution} & \colhead{Instrument} \\
                                          & \colhead{day} & \colhead{seconds}   & \colhead{$\AA$} &  & 
 }
 \startdata
\input{"./table_spec.tex"}
 \enddata
 \end{deluxetable*}

\end{document}

%% file: table_phot.tex
57496.28 & g & $>$21.03 & \nodata \\
57520.32 & g & 17.5     & 0.1  \\
57529.69 & B & 16.82 & 0.036 \\
57529.69 & U & 16.64 & 0.028 \\
57529.69 & UVM2 & 16.96 & 0.036 \\
57529.69 & UVW1 & 16.84 & 0.036 \\
57529.69 & UVW2 & 17.44 & 0.036 \\
57529.69 & V & 16.94 & 0.071 \\
57531.68 & B & 16.78 & 0.054 \\
57531.68 & U & 16.63 & 0.036 \\
57531.69 & UVM2 & 16.95 & 0.032 \\
57531.68 & UVW1 & 16.88 & 0.042 \\
57531.68 & UVW2 & 17.49 & 0.036 \\
57531.69 & V & 16.89 & 0.091 \\
57533.13 & B & 16.57 & 0.063 \\
57533.13 & U & 16.45 & 0.045 \\
57533.20 & UVW1 & 16.83 & 0.050 \\
57533.13 & UVW2 & 17.55 & 0.050 \\
57533.13 & V & 16.74 & 0.11 \\
57533.15 & UVW2 & 17.44 & 0.042 \\
57537.46 & B & 16.62 & 0.045 \\
57537.46 & U & 16.40 & 0.036 \\
57537.46 & UVM2 & 17.06 & 0.036 \\
57537.46 & UVW1 & 16.85 & 0.036 \\
57537.46 & UVW2 & 17.55 & 0.036 \\
57537.46 & V & 16.6 & 0.071 \\
57541.45 & B & 16.47 & 0.045 \\
57541.44 & U & 16.28 & 0.036 \\
57541.45 & UVM2 & 17.09 & 0.036 \\
57541.44 & UVW1 & 16.85 & 0.036 \\
57541.45 & UVW2 & 17.64 & 0.036 \\
57541.45 & V & 16.51 & 0.071 \\
57543.30 & B & 16.39 & 0.045 \\
57543.31 & U & 16.33 & 0.036 \\
57543.38 & UVM2 & 17.22 & 0.042 \\
57543.37 & UVW1 & 16.93 & 0.042 \\
57543.37 & UVW2 & 17.7 & 0.042 \\
57543.31 & V & 16.43 & 0.071 \\
57545.30 & B & 16.39 & 0.045 \\
57545.23 & U & 16.35 & 0.036 \\
57545.24 & UVM2 & 17.35 & 0.036 \\
57545.23 & UVW1 & 17.05 & 0.042 \\
57545.24 & UVW2 & 17.88 & 0.036 \\
57545.24 & V & 16.56 & 0.071 \\
57547.31 & B & 16.38 & 0.036 \\
57547.31 & U & 16.33 & 0.028 \\
57547.32 & UVM2 & 17.36 & 0.036 \\
57547.31 & UVW1 & 17.1 & 0.036 \\
57547.31 & UVW2 & 17.84 & 0.042 \\
57547.32 & V & 16.45 & 0.061 \\
57550.56 & B & 16.37 & 0.036 \\
57550.55 & U & 16.36 & 0.036 \\
57550.56 & UVM2 & 17.51 & 0.036 \\
57550.55 & UVW1 & 17.16 & 0.036 \\
57550.56 & UVW2 & 17.97 & 0.042 \\
57550.56 & V & 16.51 & 0.071 \\
57553.28 & B & 16.44 & 0.063 \\
57553.28 & U & 16.54 & 0.054 \\
57556.88 & B & 16.46 & 0.045 \\
57556.87 & U & 16.64 & 0.036 \\
57556.87 & UVM2 & 17.91 & 0.042 \\
57556.86 & UVW1 & 17.64 & 0.050 \\
57556.87 & UVW2 & 18.36 & 0.050 \\
57556.87 & V & 16.50 & 0.071 \\
57559.32 & B & 16.42 & 0.045 \\
57559.32 & U & 16.69 & 0.036 \\
57559.39 & UVM2 & 18.15 & 0.050 \\
57559.39 & UVW2 & 18.51 & 0.050 \\
57559.39 & V & 16.41 & 0.081 \\
57562.12 & B & 16.60 & 0.045 \\
57562.18 & U & 16.80 & 0.045 \\
57562.12 & UVM2 & 18.31 & 0.050 \\
57562.18 & UVW1 & 17.94 & 0.058 \\
57562.12 & UVW2 & 18.91 & 0.050 \\
57562.12 & V & 16.60 & 0.071 \\

%% file: table_HST.tex
2016-06-02 05:22:47 & 2  & COS/FUV$^a$   & G140L & 1118 - 2251  & 1500-2900 & TIME-TAG \\
2016-06-02 10:06:12 & 1  & STIS/NUV      & G230L & 1570 - 3180  & 500-1010  & NUV-MAMA \\ 
2016-06-14 04:47:38 & 1  & STIS/NUV      & G230L & 1570 - 3180  & 500-1010  & NUV-MAMA \\
2016-06-30 03:39:54 & 1  & STIS/NUV      & G230L & 1570 - 3180  & 500-1010  & NUV-MAMA \\

%% file: table_spec.tex
2016-05-27 &  57535.33& 900    & 3000-9500 &  1434 at 7500$\AA$ & P200/DBSP \\
2016-05-30 &  57538.25&1200     &  11530-13517 & 3074 at 11530$\AA$  & Keck-I/MOSFIRE \\
2016-05-30 &  57538.27&1200     &  14646-17860 & 3291 at 14646$\AA$  & Keck-I/MOSFIRE \\
2016-05-31 &  57539.17& 1200   & 3000-9500 &  1434 at 7500$\AA$     & P200/DBSP \\
2016-06-02 & 57541.35 & 4889 & 1118-2251 & 1500 at 1118$\AA$        & COS/FUV/Seg-A/G140L \\
2016-06-02 & 57541.45 & 2327 & 1570-3180 & 1010 at 3180$\AA$        & STIS/NUV-MAMA/G230L \\
2016-06-07 & 57546.34 & 220    & 3000-10000 &  1339 at 7500$\AA$    & Keck-I/LRIS    \\
2016-06-14 & 57553.21 & 2327 & 1570-3180 & 1010 at 3180$\AA$        & STIS/NUV-MAMA/G230L \\
2016-06-30 & 57569.12  & 2327 & 1570-3180 & 1010 at 3180$\AA$        & STIS/NUV-MAMA/G230L \\